\pdfoutput=1
\documentclass[a4paper,english,aps,showpacs]{revtex4}
\usepackage[T1]{fontenc}
\usepackage[latin9]{inputenc}
\usepackage{color}
\usepackage{babel}

\usepackage{array}
\usepackage{longtable}
\usepackage{float}
\usepackage{booktabs}
\usepackage{amsmath}
\usepackage{graphicx}
\usepackage{amssymb}
\usepackage{esint}
\usepackage[unicode=true, 
 bookmarks=true,bookmarksnumbered=false,bookmarksopen=false,
 breaklinks=false,pdfborder={0 0 0},backref=false,colorlinks=false]
 {hyperref}
\hypersetup{pdftitle={Planar Signatures in WMAP},
 pdfauthor={Thiago S. Pereira, L. Raul Abramo & Armando Bernui},
 pdfsubject={cosmology}}

\providecommand{\tabularnewline}{\\}

\begin{document}

\title{Searching for planar signatures in WMAP}

\author{L. Raul Abramo}

\email{abramo@fma.if.usp.br}

\affiliation{Instituto de F\'isica, Universidade de S\~ao Paulo, CP 66318, 05315-970,
S\~ao Paulo, Brazil}

\author{Armando Bernui}

\email{bernui@das.inpe.br}

\affiliation{Instituto Nacional de Pesquisas Espaciais \\
\mbox{Div. de Astrof\'{\i}sica, Av. dos Astronautas 1758, 12227-010  
S\~ao Jos\'e dos Campos -- SP, Brazil} \\
            and \\
Centro Brasileiro de Pesquisas F\'{\i}sicas \\
\mbox{Rua Dr.\ Xavier Sigaud 150, 22290-180  Rio de Janeiro -- RJ, Brazil}}

\author{Thiago S. Pereira}

\email{thiago@ift.unesp.br}

\affiliation{Instituto de F\'isica Te\'orica, UNESP - Universidade Estadual Paulista, 
Caixa Postal 70532-2, 01156-970, S\~ao Paulo, SP, Brazil}

\begin{abstract}
We search for planar deviations of statistical isotropy in the Wilkinson Microwave 
Anisotropy Probe (WMAP) data by applying a recently introduced angular-planar 
statistics both to full-sky and to masked temperature maps, including
in our analysis the effect of the residual foreground contamination and 
systematics in the foreground removing process as sources of error. 
We confirm earlier findings that full-sky maps exhibit anomalies at the 
planar ($l$) and angular ($\ell$) scales $(l,\ell)=(2,5),(4,7),$ and $(6,8)$, 
which seem to be due to unremoved foregrounds since this features are present in the 
full-sky map but not in the masked maps. On the other hand, our test detects slightly anomalous 
results at the scales $(l,\ell)=(10,8)$ and $(2,9)$ in the masked maps but not in the 
full-sky one, indicating that the foreground cleaning procedure (used to generate the full-sky map) 
could not only be creating false anomalies but also hiding existing ones. 
We also find a significant trace of an anomaly in the full-sky map at the 
scale $(l,\ell)=(10,5)$, which is still present when we consider galactic cuts of 
$18.3\%$ and $28.4\%$. As regards the quadrupole ($\ell=2$), we find a coherent 
over-modulation over the whole celestial sphere, for all full-sky and cut-sky maps.
Overall, our results seem to indicate that current CMB maps derived from WMAP
data do not show significant signs of anisotropies, as measured by our
angular-planar estimator. However, we have detected a curious coherence of
planar modulations at angular scales of the order of the galaxy's plane, which
may be an indication of residual contaminations in the full- and cut-sky maps.
\end{abstract}


\maketitle

\section{introduction\label{sec:intro}}

The statistical concordance model of the universe has now been established
with unprecedent accuracy by the Wilkinson Microwave Anisotropy Probe (WMAP). 
The five year dataset release of the WMAP team~\citep{Hinshaw:2008kr,Komatsu:2008hk} 
has shown that -- to a very good extent -- the temperature field of
the cosmic microwave background radiation (CMB) obeys a Gaussian statistic
with zero mean and a variance which is the same in every direction
in the sky. At the same time, several different teams 
\citep{deOliveiraCosta:2003pu,Eriksen:2003db,Schwarz:2004gk,Copi:2005ff,Land:2005ad,
Bernui:2005pz,Eriksen:2007pc,Lew:2008mq,Samal:2007nw,Samal:2008nv,Bernui:2008cr} 
have reported the detection of statistical peculiarities in this field,
mainly in the largest cosmological scales, which range from slightly to extremely 
unlikely within the framework of a Gaussian and statistically isotropic
universe. Among these anomalies, the most conspicuous are the lack
of power in the low-$\ell$ multipole sector and in the 2-point
correlation function \cite{deOliveiraCosta:2003pu,Copi:2005ff},
the alignment of the quadrupole $\ell=2$ and the octopole $\ell=3$
\cite{Schwarz:2004gk,Copi:2005ff,Gruppuso:2009ee}, and the so-called north-south
asymmetry \cite{Eriksen:2003db,Bernui:2005pz,Eriksen:2007pc,Bernui:2006ft,Ayaita:2009xm,
Bernui:2007eu,Bernui:2008cr,Samal:2007nw,Samal:2008nv,Pietrobon:2008ve,Pietrobon:2009qg}. 
Several attempts to explain away these statistical features in
terms of known sources of microwave radiation or peculiarities of the instrument
have been proposed, but have not yet produced a compelling explanation.

If we put aside the hypotheses that these anomalies may possibly be due to some 
residual galactic contamination~\citep{deOliveiraCosta:2006zj,Abramo:2006hs,Chiang:2006fe,Bernui:2008ei}
or even to a systematic data analysis effect~\citep{Bunn:2006nh,Li:2009zy,Naselsky:2007gt}, 
we inevitably end up reverse-engineering physical and/or astrophysical phenomena
in order to explain these effects. Although the former explanations
are important and deserve attention on their own, the possibility
that a new mechanism may explain some of the unknown CMB features
is still far from being ruled out. 
In such a case, we have essentially two different approaches: a theoretical 
(bottom-up) and a phenomenological (top-down) one. 

In the bottom-up approach, models for the evolution of the universe are 
formulated through physical principles in such a way as to account for 
deviations of Gaussianity and/or statistical isotropy (SI). 
These models usually invoke the existence of non-trivial cosmic 
topologies~\citep{Luminet:2003dx,Riazuelo:2003ud,HipolitoRicaldi:2005eh}, 
primordial magnetic fields \citep{Kahniashvili:2008sh,Kahniashvili:2008hx,Bernui:2008ve,Caprini:2009vk,Seshadri:2009sy}, 
local~\citep{Gordon:2005ai,Prunet:2004zy,Campanelli:2007qn,Campanelli:2009tk}
and global \citep{Pereira:2007yy,Pitrou:2008gk,Gumrukcuoglu:2007bx,Ackerman:2007nb}
manifestations of anisotropy and even exotic models of 
inflation~\citep{Erickcek:2008sm,Erickcek:2009at,Donoghue:2007ze,Shtanov:2009wp,Kawasaki:2008sn,Kawasaki:2008pa}, only to mention a few of the possibilities.

In the phenomenological top-down approach, we design statistical
tools to test the robustness of temperature maps against the hypotheses
of Gaussianity and SI. These approaches 
include statistics of multipole alignments using Maxwell's multipole
vectors \citep{Copi:2003kt,Copi:2005ff,Abramo:2006gw,Katz:2004nj},
constructs of \textit{a posteriori} statistics \cite{Bernui:2005pz}
and functional modifications of the two-point correlation function
\citep{Pullen:2007tu,Hajian:2003qq,Hajian:2004zn,Hajian:2005jh}. 

One interesting feature of the top-down approach is that it can shed
some light on the source of the reported anomalies without the need
to invoke any specific cosmological model, but rather acting as
a further guide for theorists. However, due to the
intrinsic generality of this approach, it is sometimes unclear how to select the
important degrees of freedom in order to construct
meaningful statistical estimators. This is in fact a natural
consequence of model-independent approaches, which at some stage force
us to rely on our theoretical prejudices about the statistical nature
of the CMB in order to construct estimators of non-Gaussianity and/or
statistical anisotropy. 

We can instead follow another route and let the construction of our
statistics be guided primarily by what the observations tell us, rather than 
by what we think they should look like.
Here we are primarily interested in answering: how to detect planar 
signatures in CMB maps?
One possibility was recently proposed in~\cite{Pereira:2009kg}, 
where the two-point correlation function is decomposed in such a way as to 
account for the presence of symmetry planes in temperature maps. 
Given that CMB experiments are confined to collecting
data from inside our galactic plane, it is conceivable that we may
find planar signatures in the data if foregrounds were improperly
removed. Besides being observationally motivated, this approach leads
naturally to both an unbiased estimator and a rotationally invariant test of 
statistical anisotropy, therefore alleviating the difficulties mentioned above.

In this work we employ the angular-planar power spectrum analysis introduced  
in~\cite{Pereira:2009kg} to search for planar signatures both in partial 
and full sky-coverage temperature maps, allowing for possible 
effects due to residual foreground contaminations. Our analysis may therefore help to 
shed some light in the origin of the reported anomalies and test the robustness of 
cleaned maps due to foreground contamination in a model-independent way.\\

This work is organized as follows: we start $\S$\ref{sec:2pacf}
by reviewing the construction of the two-point correlation function,
and we show how it can include planar deviations of SI. 
We then show that a simple chi-square analysis can be designed to detect 
such planarities in any theoretical cosmological model. We end that 
Section with a simple example of planarity which could be 
easily detected through our tests. In $\S$\ref{sec:Probabilities}
we discuss how the probabilities of measuring particular values of cosmological 
observables are usually calculated, and how this calculations can be extended to include 
the uncertainty inherent in this measurement. In Section $\S$\ref{sec:foregrounds} 
we explain how to include the effect of foreground contamination in our analyses, 
and present our results for full-sky temperature maps. After presenting the results 
of the analyses for masked maps in $\S$\ref{sec:masks}, we conclude and give some 
perspectives of further developments.

\section{Isotropic and anisotropic two-point correlation functions\label{sec:2pacf}}

The starting point of our analysis is the temperature fluctuation
field $\Delta T(\hat{\boldsymbol{n}})$. According to standard inflationary
models, this is a random field of which we have access to one single
realization. If we decompose this field in spherical harmonics:
\begin{equation}
\Delta T(\hat{\boldsymbol{n}})=\sum_{\ell,m}a_{\ell m}Y_{\ell m}(\hat{\boldsymbol{n}}) \, ,
\label{DeltaT}
\end{equation}
its randomness can be attributed to the multipolar coefficients $a_{\ell m}$,
and the statistics of the field can be characterized by constructing
the statistical moments of these coefficients. For simplicity, we 
suppose that this field is Gaussian with zero mean $\langle\Delta T(\hat{\boldsymbol{n}})\rangle=0$,
in which case the two-point correlation function,
\begin{equation}
C(\hat{\boldsymbol{n}}_{1},\hat{\boldsymbol{n}}_{2})=\langle\Delta T(\hat{\boldsymbol{n}}_{1})\Delta T(\hat{\boldsymbol{n}}_{2})\rangle=\sum_{\ell_{1},m_{1}}\sum_{\ell_{2},m_{2}}\langle a_{\ell_{1}m_{1}}a_{\ell_{2}m_{2}}^{*}\rangle Y_{\ell_{1}m_{1}}(\hat{\boldsymbol{n}}_{1})Y_{\ell_{2}m_{2}}^{*}(\hat{\boldsymbol{n}}_{2}) \, ,
\label{full-cf}
\end{equation}
is the only relevant statistical moment. Equivalently, the covariance
matrix $\langle a_{\ell_{1}m_{1}}a_{\ell_{2}m_{2}}^{*}\rangle$ encloses
all the statistical information of the multipolar coefficients in
Eq. (\ref{DeltaT}). Notice that the hypothesis of Gaussianity makes
no restriction whatsoever upon the rotational properties of the two-point
function $C(\hat{\boldsymbol{n}}_{1},\hat{\boldsymbol{n}}_{2})$.
It is therefore perfectly legitimate to restrict our considerations
to the Gaussian case when looking for deviations of SI, noting that
these two questions should be clearly stated and, as much as possible,
decoupled.

\subsection{Isotropic correlation function}

The statistical counterpart of the concordance $\Lambda\mbox{CDM}$
model predicts a universe which, in its linear regime, is not only
Gaussian but also statistically isotropic. This means that the temperature 
fluctuation field (and its variance) has no privileged directions,
and that the variance depends only on the angular separation of two 
given points in the CMB sky:
\begin{equation}
C(\hat{\boldsymbol{n}}_{1}\cdot\hat{\boldsymbol{n}}_{2})=
\sum_{\ell}\frac{2\ell+1}{4\pi}C_{\ell}P_{\ell}(\hat{\boldsymbol{n}}_{1}
\cdot\hat{\boldsymbol{n}}_{2}) \, .
\label{fc-iso}
\end{equation}

In harmonic space, rotational invariance means that the correlation 
function is completely diagonal: 
\begin{equation}
\langle a_{\ell_{1}m_{1}}a_{\ell_{2}m_{2}}^{*}\rangle=C_{\ell_1}
\delta_{\ell_1\ell_2}\delta_{m_1 m_2} \, ,
\label{cov-matrix-iso}
\end{equation}
with its eigenvalues determined by the theoretical \textit{angular
power spectrum} $C_{\ell_1}$. Together, Gaussianity and SI 
comprise a very severe set of restrictions to the temperature
field Eq. (\ref{DeltaT}), and allows no expressions other than (\ref{fc-iso})
and (\ref{cov-matrix-iso}) to specify its properties.

\subsection{Planar correlation function}

Deviations of SI presuppose functional deviations of the two-point function
from Eq. (\ref{full-cf}). Despite the infinite possibilities that are open when
one relaxes the requirement that the two-point function depends only on the separation angle
$\cos\theta=\hat{\boldsymbol{n}}_1\cdot\hat{\boldsymbol{n}}_2$, 
we should still remember that the 
unit vectors $\hat{\boldsymbol{n}}_{1}$ and $\hat{\boldsymbol{n}}_{2}$ share a 
common origin at the center of the CMB sphere (in other words, the center of
our surface of last scattering.) 
One of the possibilities that do not assume isotropy is
to consider $C=C(\hat{\boldsymbol{n}}_{1},\hat{\boldsymbol{n}}_{1})$
\cite{Pullen:2007tu}, but in this case we would not actually be measuring
correlations of two points, but rather the self-correlation of all individual 
points in the sky (i.e., the coincidence limit of the two-point correlation function.) 
Another possibility is to consider 
$C=C(\hat{\boldsymbol{n}}_{1},\hat{\boldsymbol{n}}_{2})$
as given by expression (\ref{full-cf}), with the two vectors completely
independent of each other \cite{Hajian:2003qq}. Unfortunately this
function is too general, and leads to rampant arbitrariness in the construction
of statistical estimators of anisotropies, as we mentioned in $\S$\ref{sec:intro}. \\

A third possibility was pointed in ~\cite{Pereira:2009kg}: 
for any pair of unit vectors in the CMB sphere,
not only their angular separation is well-defined, but also the plane
where they stay. That is, since these vectors are restricted to have
a common origin, both their scalar and cross products are uniquely defined.
This suggests that we incorporate this vectorial dependence into the
two-point correlation function:
\begin{equation}
C(\hat{\boldsymbol{n}}_{1},\hat{\boldsymbol{n}}_{2})=C(\hat{\boldsymbol{n}}_{1}\times\hat{\boldsymbol{n}}_{2}) \, ,
\label{fc-aniso-formal}
\end{equation}
which can now account for both angular and planar modulations of the
temperature field. Note that now vectors with the same angular
separation are not necessarily correlated in the same way. This feature is what
makes the functional form above suitable for detecting planar signatures in
CMB sky maps, no matter if they have an astrophysical origin (as in, e.g., the 
galactic plane) or if they are artificially imprinted on the maps (e.g., poor masks, 
inhomogeneous observation function, etc.)

In practice, just as in the case of the usual (isotropic) $C_\ell$'s, 
it is easier to work with (\ref{fc-aniso-formal})
in harmonic space. If we introduce the definition $\mathbf{n}\equiv\hat{\boldsymbol{n}}_{1}\times\hat{\boldsymbol{n}}_{2}=
\{\sin\vartheta,\theta,\phi\}$,
where $\vartheta=\arccos\hat{\boldsymbol{n}}_{1}\cdot\hat{\boldsymbol{n}}_{2}$,
we obtain the following decomposition:

\begin{equation}
C(\mathbf{n})=\sum_{\ell}\sum_{l,m}\frac{2\ell+1}{\sqrt{4\pi}}\mathcal{C}_{\ell}^{lm}
P_{\ell}(\cos\vartheta)Y_{lm}(\hat{\mathbf{n}}) \, , \quad l\in2\mathbb{N} \, ,
\label{fc-aniso}
\end{equation}
where the restriction of the sum above to even values of $l$ is
a consequence of the reciprocity relation $C(\hat{\boldsymbol{n}}_{1},\hat{\boldsymbol{n}}_{2})=
C(\hat{\boldsymbol{n}}_{2},\hat{\boldsymbol{n}}_{1})$. The decomposition above is 
equivalent to considering modulations of the function $C_{\ell}(\hat{\mathbf{n}})$ in 
the sphere, where the usual angular power spectrum $C_{\ell}$ is now given by the 
monopole $\mathcal{C}_{\ell}^{00}$. In fact, we could have started with this
modulations rather than (\ref{fc-aniso}), but in such a case we would 
perhaps neglect the geometrical interpretation of the normal vector
$\hat{\mathbf{n}}$, which is crucial in our analysis.

By equating Eqs. (\ref{fc-aniso}) and (\ref{full-cf}), it is possible to show~\cite{Pereira:2009kg} 
that the coefficients $\mathcal{C}^{lm}_\ell$ are 
directly related to the temperature multipolar coefficients $a_{\ell m}$. We have that:

\begin{equation}
\frac{\mathcal{C}_{\ell}^{lm}}{\sqrt{2l+1}}=2\pi\sum_{\ell_{1},m_{1}}\sum_{\ell_{2},m_{2}}
\langle a_{\ell_{1}m_{2}}a_{\ell_{2}m_{2}}\rangle\left(\begin{array}{ccc}
l & \ell_{1} & \ell_{2}\\
m & m_{1} & m_{2}\end{array}\right)I_{\ell_{1}\ell_{2}}^{l,\ell} \, ,
\label{Clmell}
\end{equation}
where
\begin{equation}
I_{\ell_{1}\ell_{2}}^{l,\ell}\equiv\sum_{m}(-1)^{m}\lambda_{\ell_{1}m}\lambda_{\ell_{2}m}
\int_0^\pi d(-\cos\vartheta)\, P_{\ell}(\cos\vartheta)e^{im\vartheta}\left(\begin{array}{ccc}
l & \ell_{1} & \ell_{2}\\
0 & m & -m\end{array}\right) \, ,
\label{Int-l-ell}
\end{equation}
with $\lambda_{\ell m}$ a set of real coefficients resulting
from the $\vartheta$ integration and which are zero unless $\ell+m=$
even (see the Appendix of~\cite{Pereira:2009kg} for more details). These expressions
show, first, that the angular-planar multipolar coefficients $\mathcal{C}_{\ell}^{lm}$
can be calculated from first principles in any theoretical model which predicts a specific
form for the covariance matrix $\langle a_{\ell_{1}m_{1}}a_{\ell_{2}m_{2}}^{*}\rangle$
like, for instance, models of anisotropic inflation 
\citep{Pereira:2007yy,Pitrou:2008gk,Gumrukcuoglu:2007bx,Ackerman:2007nb}.
And second, given any observed temperature map, we can estimate the values 
of all ${\mathcal{C}}^{lm}_\ell$'s by computing the sum in Eq. (\ref{Clmell}) with
the actual coefficients $a_{\ell m}$ -- see the next Subsection.
Note also that the coefficients (\ref{Clmell}) are not restricted
to temperature maps, and can be equally applied in the analysis of
CMB polarization, large-scale structure maps or in fact any map on $S^2$.

\subsection{Statistical estimators of anisotropy}

The multipolar angular-planar coefficients (\ref{Clmell}) were defined
in terms of an ensemble average of the temperature multipolar coefficients
$a_{\ell m}$. Of course, we only observe one universe, which
means that the cosmic variance in the determination of the coefficients
$\mathcal{C}_{\ell}^{lm}$ is a severe restriction that we have to
live with. This means that, if we want to evaluate the statistical properties of these 
coefficients, like mean values or variances, we have to build functions which can at best 
estimate these quantities. One obvious possibility is the following unbiased estimator:
\begin{equation}
\mathcal{C}_{\ell}^{lm}\rightarrow2\pi\sqrt{2l+1}\sum_{\ell_{1},m_{1}}
\sum_{\ell_{2},m_{2}}a_{\ell_{1}m_{2}}a_{\ell_{2}m_{2}}\left(\begin{array}{ccc}
l & \ell_{1} & \ell_{2}\\
m & m_{1} & m_{2}\end{array}\right)I_{\ell_{1}\ell_{2}}^{l,\ell} \, ,
\label{Clmell-hat}
\end{equation}
which can be uniquely calculated for a given temperature map. 

Following this prescription, we can now estimate the statistical moments
of (\ref{Clmell-hat}). In particular, its mean value was already calculated, and is 
given by $\sqrt{2l+1}$ times the right-hand side of (\ref{Clmell}). Notice that, 
under the hypothesis of SI, it follows from Eqs. (\ref{Clmell-hat}) and 
(\ref{cov-matrix-iso}) that~\cite{Pereira:2009kg}:
\begin{equation}
\langle\mathcal{C}_{\ell}^{lm}\rangle_{\rm SI}=C_{\ell}\delta_{l0}\delta_{m0} \, .
\label{SI}
\end{equation}
Conversely, if the only non-zero $\mathcal{C}_{\ell}^{lm}$'s are given
by $l=m=0$, then $\mathcal{C}_{\ell}^{00}=C_{\ell}$, and we conclude
that SI is achieved if and only if the coefficients $\mathcal{C}_{\ell}^{lm}$'s 
are of the form (\ref{SI}). From now on, we will restrict our analysis to $l\neq0$, 
since this is the first non-trivial case of planarity.

\subsection{A $\chi^2$ test of statistical isotropy}

We have determined above, in a statistically isotropic universe,  that all 
$\mathcal{C}_{\ell}^{lm}$'s with $l \neq 0$ are 
random variables with zero mean. We can now ask, with a set of
$\mathcal{C}_{\ell}^{lm}$'s at hand, how do we check whether they respect
statistical isotropy -- and in case they do not, which type of anisotropy they 
correspond to? 

If we have a set of zero-mean random variables as in (\ref{SI}), an obvious choice 
would be to construct the associated chi-square test in the same way as the 
usual $C_\ell$'s are constructed as a chi-square fit from the $a_{\ell m}$'s. However, 
there is nothing which prevents us from applying this test to models where (\ref{SI}) 
{\it does not} hold -- all that is needed is to subtract from the $\mathcal{C}_{\ell}^{lm}$'s 
its expectation value given by a fiducial model. Hence we can define the following 
(reduced) chi-square test:
\begin{equation}
\label{chi2_0}
(\chi^2_\nu)^l_\ell \equiv \frac{1}{2\ell+1}
\sum_{m=-\ell}^\ell 
\frac{\left| \mathcal{C}_{\ell}^{lm} -  \langle \mathcal{C}_{\ell}^{lm} \rangle \right|^2}
{(\sigma^{lm}_\ell)^2} \; ,
\end{equation}
where $\langle{\mathcal{C}}_{\ell}^{lm}\rangle$ and $(\sigma^{lm}_\ell)^2$ 
are respectively the expectation value and the variance of the $\mathcal{C}_{\ell}^{lm}$'s in 
some particular model. Notice that, since we are summing over $m$ in 
the definition of the $(\chi^2_\nu)^l_\ell$, our test is rotationally invariant. This 
is a crucial property, since we are pursuing a ``blind test'' of anisotropy where we do 
not know what type of directionality we are looking for.\\

Now that we have defined a general test of anisotropy, we need a model to be tested. As 
it turns out, the most successful model we have is the concordance $\Lambda$CDM model, 
for which SI holds. Under the hypothesis of SI, not only $\langle {\mathcal{C}}_{\ell}^{lm} \rangle$ can be easily calculated, but also the variance $(\sigma^{lm}_\ell)^2$ have a simple, $m$-independent, expression:
\begin{equation}
\label{sigma2}
(\sigma^{lm}_\ell)^2_{\rm SI} \equiv (\sigma^l_\ell)^2 =
8 \pi^2 \sum_{\ell_1, \ell_2} C_{\ell_1} C_{\ell_2} 
\left( I^{l,\ell}_{\ell_1 , \ell_2} \right)^2 \; .
\end{equation}

For this model, it is clear from Eqs. (\ref{SI}), (\ref{chi2_0}), and (\ref{sigma2}) 
that we will have $\langle (\chi^2_\nu)^l_\ell \rangle = 1$ as long as $l\neq0$, 
so we define the angular-planar $\chi^2$ measure of the deviations of any 
map from SI to be:
\begin{equation}
\label{chi2}
(\bar{\chi}^2)^l_\ell \equiv (\chi^2_\nu)^l_\ell - 1 \; .
\end{equation}
If the test $\bar\chi^2$ is significantly positive or negative (allowing for
some variance which can be different for different maps), this will be an indication 
of deviations from SI.

\subsection{\label{sec:exemplo} An example of planarity}

Before we move on to apply (\ref{chi2}) to actual CMB data, it is important to have 
some intuition on the type of planarity which could be revealed by this test. 
We can ask, for example, what pattern of modulations would a Gaussian and SI
temperature map $\Delta T(\hat{\boldsymbol{n}})$ inherit when we
apply to it a window function $W(\hat{\boldsymbol{n}})$. Depending
on the shape of this function, the resulting map
\begin{equation}
\Delta\overline{T}(\hat{\boldsymbol{n}})=
\Delta T(\hat{\boldsymbol{n}})W(\hat{\boldsymbol{n}}) \; ,
\label{DT-barra}
\end{equation}
will contain planar modulations which our test (\ref{chi2}) can easily detect. 
Notice, however, that this procedure is not the same as `cutting' or making a 
`hole' in the sphere $S^2$ in order to neglect the contaminations from our galaxy 
with the help of a mask (as we will do in \S\ref{sec:masks}). 
The effect of the window function above is conceptually simpler, 
and corresponds to a region with enhanced/suppressed fluctuations,
which naturally induces strong anisotropic correlations on $\Delta\overline T$. In 
particular, the resulting angular power spectrum $\overline C_\ell$ will 
not be trivially related to the isotropic $C_\ell$'s, but rather it will be given by a 
convolution with the latter which couples different angular scales through 
some kernel \cite{Prunet:2004zy,Hivon:2001jp}. Consequently, 
even the angular spectrum $\overline C_\ell$ would hint strongly towards anisotropy, 
although in this case no further information of planarity could be extracted from it.
\begin{figure}[H]
\begin{center}
\includegraphics[scale=0.29]{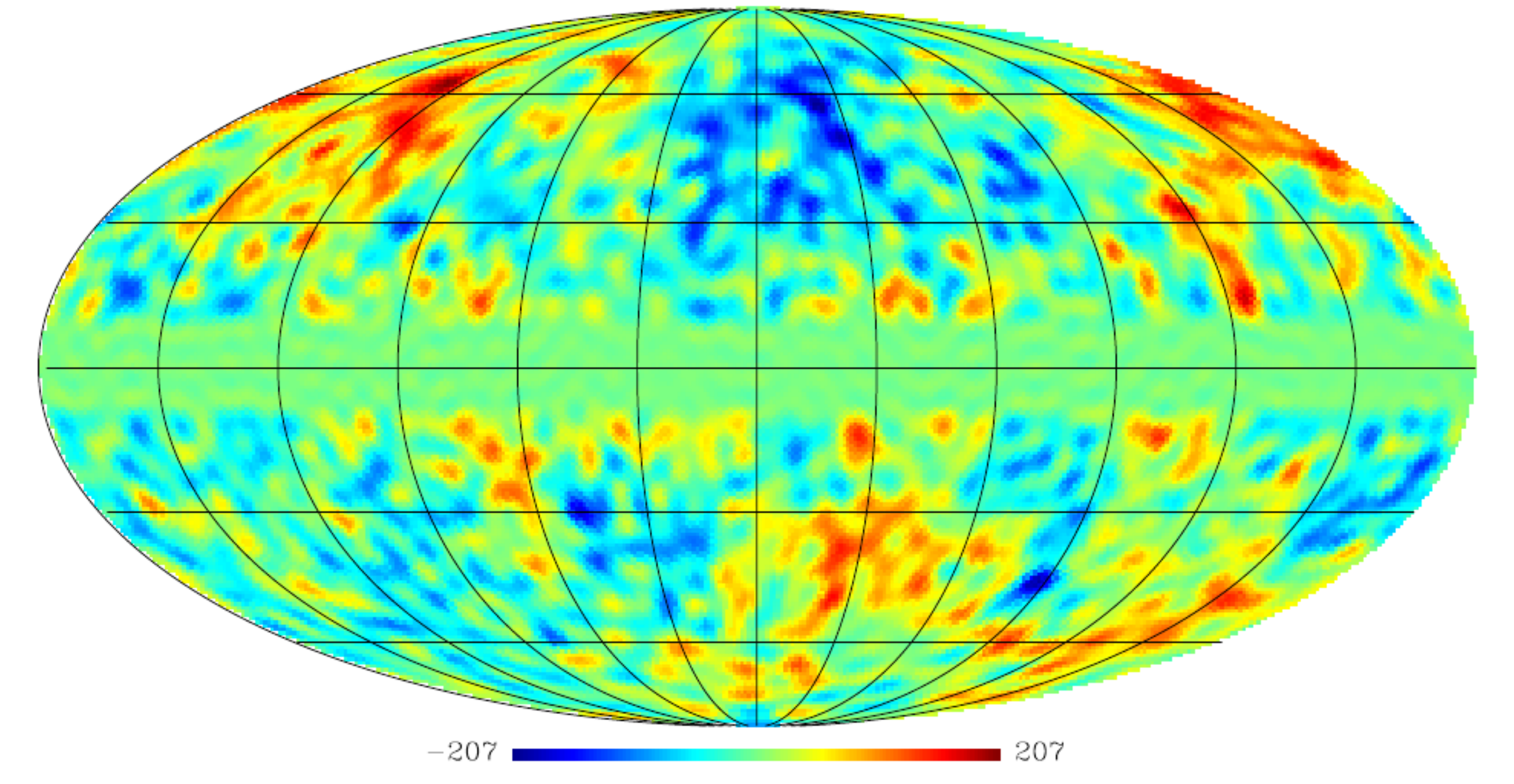}
\caption{Temperature map with a planar signature around the equator.
See the text for more details.}
\label{mapa50}
\end{center}
\end{figure}
Having these distinctions in mind, let us take as an example a simple window 
function which gives zero weight for fluctuations around an equatorial strip of 
fixed width $\Delta\theta$, i.e.:
\begin{equation}
W(\theta)=\begin{cases}
0 & \theta\in\frac{1}{2}[\pi\pm\Delta\theta]\\
1 & \mbox{otherwise}\,.\end{cases}
\label{window-function}
\end{equation}
This window function introduces a disk around the equator where all
points have the same temperature, as can be seen in Fig. (\ref{mapa50}). 
If we further decompose the function (\ref{window-function}) in spherical harmonics and use 
(\ref{DeltaT}), we arrive at \cite{Prunet:2004zy}:
\begin{equation}
\overline{a}_{\ell m}=\sum_{\ell',m'}K^{\ell'm'}_{\ell m}a_{\ell' m'}\,,
\label{alm-barra}
\end{equation}
with $K^{\ell'm'}_{\ell m}$ given by
\[
K^{\ell'm'}_{\ell m}=\sum_{L,M}w_{\scriptsize{LM}}\int d^2\hat{\boldsymbol{n}}\;
Y_{LM}(\hat{\boldsymbol{n}})Y_{\ell'm'}(\hat{\boldsymbol{n}})Y_{\ell m}^*(\hat{\boldsymbol{n}})
\]
and where $w_{LM}$ are the coefficients resulting from the decomposition of (\ref{window-function}) into spherical harmonics.

Although, in principle, expression (\ref{alm-barra}) could be used to explicitly evaluate 
both (\ref{sigma2}) and (\ref{chi2}), the resulting expression is not clarifying. In practice 
it is much easier to obtain the same result by simulating these maps and applying the 
$\bar\chi^2$ test to each one of them. Nonetheless, the relevance of Eq. 
(\ref{alm-barra}) lies in the linear relation that is established between 
$\overline{a}_{\ell m}$ and $a_{\ell m}$, which shows that the resulting 
temperature field $\Delta\overline{T}$ will still be Gaussian, provided that 
$\Delta T$ is Gaussian. In other words, in this example we have a Gaussian but
anisotropic field ($\Delta\overline{T}$), and therefore we will be performing a 
test of SI only, which is completely decoupled from the issue of Gaussianity. \\

\begin{figure}[H]
\begin{center}
\includegraphics[scale=0.67]{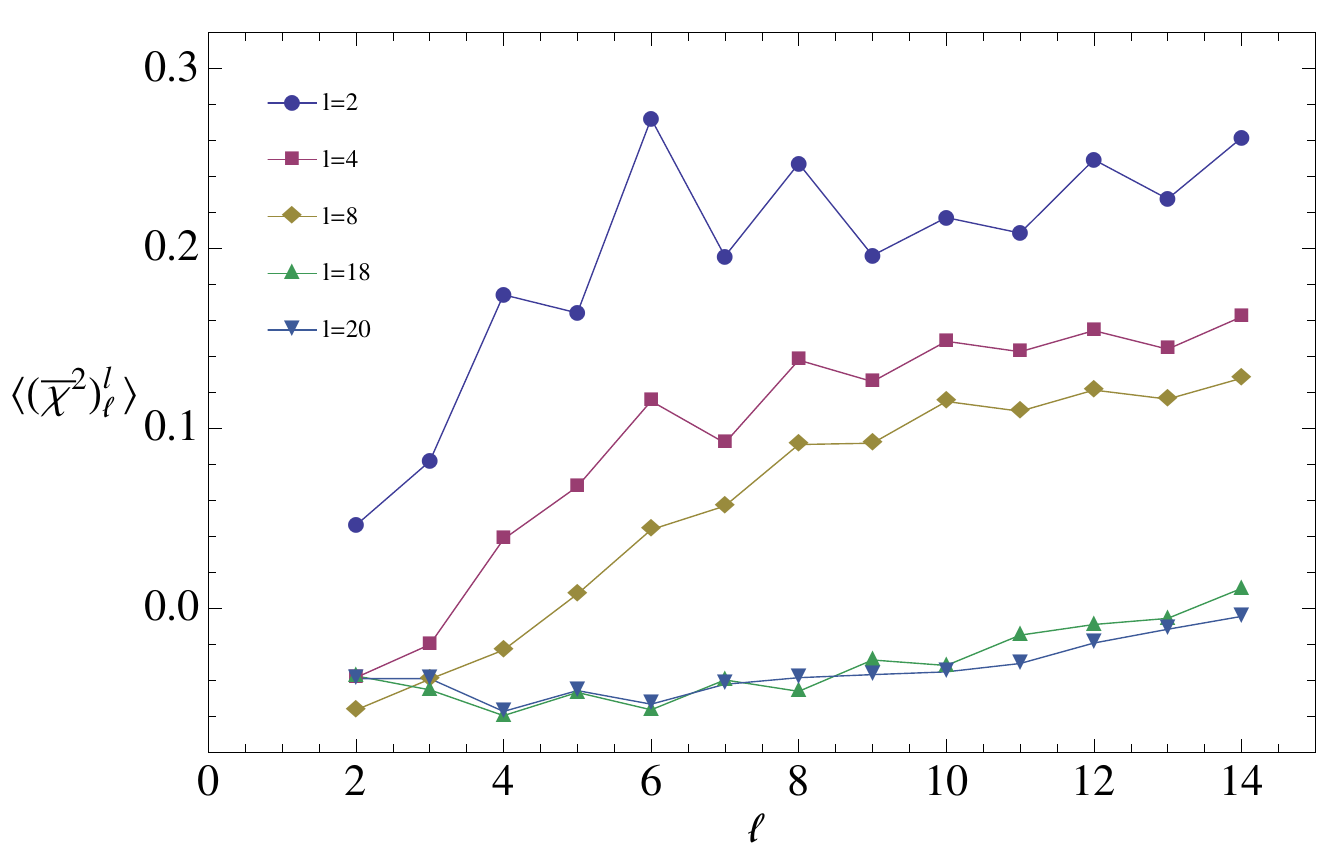}
\caption{The angular-planar $\bar\chi^2$ test averaged over $2\times 10^3$ maps with a planar modulation of $\pm 25^\circ$ around the equator. The figure shows 
$\langle(\bar{\chi}^2)^l_\ell\rangle$ as a function of $\ell$ for some particular scales in the range $l\in[2,20]$.}
\label{blell-planar}
\end{center}
\end{figure}

Following the above prescription, we have applied the estimator (\ref{chi2}) 
to $2\times 10^3$ maps of the form (\ref{alm-barra}), with the window function 
chosen to cover an equatorial strip $50^\circ$ wide; the result is shown in 
Fig.~(\ref{blell-planar}). The first thing we notice in this figure is that 
small planar scales (i.e., $\alpha\propto l^{-1}\ll1$, where $\alpha$ is the 
angle formed by the normal of the planes that lie inside the disk and the $z$ axis) 
have a roughly constant planar modulation for all the angular scales (reflected
in $\ell$) we tested. This is indeed what we would expect, for in this case we are probing 
the temperatures at points lying in circles which are nearly parallel to the equator where, 
by construction, the temperature is constant. 

As we examine larger and larger planar scales (smaller $l$'s), we see that the 
modulations become constant only below a certain angular threshold, 
corresponding to those vectors that still live inside the constant temperature 
strip. Hence, for the planar scales $l=2$, 4 and 8, the 
angular correlations level out at angular 
scales of approximately $\ell \simeq 6$, 10 and 12, respectively. 
Notice that the plateau of the $\bar\chi^2$ for each $l$, at $\ell \simeq$ 10 - 12, becomes
smaller as $l$ grows. This reflects the fact that the $50^\circ$ disk induces 
correlations on large {\it planar} scales. These correlations result in a stronger 
signal in $\bar\chi^2$ for the small {\it angular} 
scales ($\ell \gg 1$) simply because for smaller angular separations 
there are more pairs of points that contribute coherently to $\bar\chi^2$ within the disk.

For a statistically isotropic map we would expect the $(\bar\chi^2)^l_\ell$ to be randomly distributed around zero. But in the example 
above we see that this does not happen: there is a coherence of the $(\bar\chi^2)^l_\ell$ 
over many $l$'s and $\ell$'s, denoting the preferred directions/planes.
Hence, the main lesson from this example is that angular-planar features
pointing towards statistical anisotropy can sometimes be individual anomalous 
values of some $(\bar\chi^2)^l_\ell$'s, which would correspond to angular-planar
anomalies at particular angular scales $\ell$ and planar scales $l$. However,
we can also have scenarios where the individual $(\bar\chi^2)^l_\ell$'s 
are not particularly anomalous, but the coherence of the test over a range of 
$\ell$'s and $l$'s is an indication of some preferred regions or directions.
The first situation [an individual anomalous $(\bar\chi^2)^l_\ell$] points
towards a sharp anisotropic feature, whereas the second scenario [our present
example, of a coherent feature in all the $(\bar\chi^2)^l_\ell$'s] points
towards a wider region of anomalous correlations.

To conclude this example, we mention that although it was artificially built to 
illustrate the use of the angular-planar estimator, we will see in 
\S\ref{sec:masks} that similar modulations are found in cleaned CMB maps, even 
after applying the more realistic masks KQ75 and KQ85.

\section{\label{sec:Probabilities} Calculating Probabilities}

Given that our main goal is to determine how ``typical'' our universe is according
to the test (\ref{chi2}), we will now revisit the question of the probabilities 
for such tests in cosmology. \\

Let us suppose that we want to calculate the probability of detecting an observable 
inside a specific range of values, {\it according to a given cosmological model}. 
For instance, we may genuinely ask what is the probability that in a ``randomly selected''
Gaussian and SI universe we would measure one of the 
$(\bar{\chi}^2)^l_\ell$ with a value equal or smaller (greater) than the value which 
is actually measured. Often, this question is analyzed as follows:
suppose that the probability density function (pdf) of a given observable 
$x$ in this particular model/theory is known and given by $\mathcal{P}_{\scriptsize{\rm th}}$. 
Let us suppose further that the observation of this quantity
gave us a value $x_0$. Then, the number:
\begin{equation}
P_{\leq}
\equiv
\int^{x_0}_{-\infty}\mathcal{P}_{\scriptsize{\rm th}}(x)\, dx 
\label{P>0}
\end{equation}
denotes the probability of detecting a value $x$ equal to or smaller
than $x_0$, {\it according to this model}. The probability of having 
$x > x_0$ is then given by $P_{>}=1-P_{\leq}$.
However, these probabilities assume that we have measured the value $x_0$ with
infinite precision, so they say nothing about an important question
that we must deal with: typically, the measurement of $x_0$ has itself an 
uncertainty, which should be folded into the final probability that the 
observations match the theoretical expectations.

Since no measurement (of the CMB or any other physical observable) 
will give us a result with infinite precision, our measurements should also be 
regarded as random events. Therefore, in a more rigorous approach,
we would have to consider $x_0$ itself as a random variable.
In the case of CMB, however, this would be only part of the whole picture, 
since the randomness of the measurements of $x_0$ should also be related 
to the way that this data is reduced to its final form. This happens because 
different map cleaning procedures will lead to different values 
for $x_0$. This difference induces a variance in the data which 
reflects the remaining foreground contamination of the temperature maps. We will 
come back to this point later, after we show how to calculate probabilities 
when both $x$ and $x_0$ are random variables.

\subsection{Difference of random variables}

Suppose that $x$ and $y$ are two random variables, uniformly distributed
over the range $[0,1]$. What is the pdf of the new variable
$z=x-y$? Intuitively, we would expect this pdf to be zero for $z=\pm1$,
since the pairs of events $(x,y)=(1,0)$ and $(0,1)$ are the least frequent
ones. By the same reasoning, we would also expect a peak around zero.
The correct answer is a triangular pdf for $z$:
\[
\mathcal{P}(z)= 1-|z| \; ,
\]
which can be calculated by convolving the pdf's of $x$ and $y$.
This result can be generalized to include any type of continuous random
variables \cite{GutAllan}. This is precisely what we need, and if
we define:
\[
x \equiv (\bar{\chi}^2)_{\ell(\scriptsize{\mbox{th}})}^{l} \, ,
\quad 
y \equiv (\bar{\chi}^2)_{\ell(\scriptsize{\mbox{obs}})}^{l} \, ,
\quad 
z \equiv x-y \, , 
\]
then the pdf of $z$ will be formally given by
\begin{equation}
\mathcal{P}(z)\equiv
\left(\mathcal{P}_{\scriptsize{\mbox{obs}}}*\mathcal{P}_{\scriptsize{\mbox{th}}}\right)(z)
=\int_{-\infty}^{\infty}\mathcal{P}_{\scriptsize{\mbox{obs}}}(y)
\mathcal{P}_{\scriptsize{\mbox{th}}}(z+y) \, dy \,
=\int_{-\infty}^{\infty}\mathcal{P}_{\scriptsize{\mbox{obs}}}(x-z)
\mathcal{P}_{\scriptsize{\mbox{th}}}(x) \, dx \, . 
\label{pdf-conv}
\end{equation}
Notice that this probability density will be automatically normalized
if both $\mathcal{P}_{\scriptsize{\mbox{obs}}}$ and 
$\mathcal{P}_{\scriptsize{\mbox{th}}}$ are. 

According to our definitions, the probability that in a randomly selected 
$\Lambda$CDM universe we detect a value $x\leq y$ is translated into the 
probability of having $z \leq 0$, and this is simply the area under 
$\mathcal{P}(z)$ for $z\in(-\infty,0]$: 
\begin{equation}
P_{\leq}=\int^0_{-\infty}\mathcal{P}(z)\, dz \, . 
\label{P>}
\end{equation}
Moreover, in the ideal case where observations are made with infinite
precision, $\mathcal{P}_{\scriptsize{\mbox{obs}}}(y)=\delta(y-x_0)$
and we recover (\ref{P>0}). The reader must be careful, though, not
to think of (\ref{P>0}) as a lower bound to (\ref{P>}), since
the distributions $\mathcal{P}_{\scriptsize{\mbox{obs}}}$ and 
$\mathcal{P}_{\scriptsize{\mbox{th}}}$ are not necessarily symmetric about 
their mean values. In other words, a large distance from $y$ to the {\it most probable} 
theoretical value $x$ would not, by itself, constitute sufficient grounds to claim 
that this measurement of $y$ is `unusual', since its dispersion can be wide enough 
to render it `usual' according to (\ref{P>}).

Finally, there remains the question of how to obtain the probability densities
$\mathcal{P}_{\scriptsize{\mbox{obs}}}$ and $\mathcal{P}_{\scriptsize{\mbox{th}}}$.
These functions can be computed numerically,
provided that the number of realizations of the random variables $y$
and $x$ is large enough, since in this case the histograms for these variables 
can be considered as piecewise constant functions which approximate the real pdf's.
For the case of the (`theory') variable $x$ defined above we have 
run $2\times10^{4}$ Monte Carlo simulations
of Gaussian and statistically isotropic CMB maps using the $\Lambda\mbox{CDM}$
best-fit $C_{\ell}$'s provided by the WMAP team \cite{lambda}.
With these maps we have then constructed $2\times10^{4}$ realizations of
the variable $x$. This procedure was also carried using masked $C_{\ell}$'s 
(corresponding to masks KQ85 and KQ75) to construct $2\times10^4$ masked versions 
of the variable $x$.

The simulation of the variable $y$, related to the observations, is less trivial, and is
intrinsically related to the way we estimate contamination from residual foregrounds.
We will now detail this procedure.

\section{Estimating the residual foregrounds\label{sec:foregrounds}}

As mentioned before, cosmic variance is an unavoidable limitation which can be 
estimated by, e.g, running many different realizations of our theoretical random variable $x$. 
However, cosmic variance is far from being the sole source of uncertainties in CMB 
experiments. 
As is well-known, not only instrumental noise, but systematic errors (e.g., in the 
map-making process), the inhomogeneous scanning of the sky (i.e., the exposure 
function of the probe), or unremoved foreground  emissions (even after applying a 
cut-sky mask) could corrupt -- at distinct levels -- the CMB data.

Foreground contamination, on the other hand, may have several different sources, 
and consequently several different ways of being included, many of which would 
certainly go beyond the scope of the present work. However, since different teams 
apply distinct procedures on the raw data in order to produce a final map, we will 
make the hypothesis that maps cleaned by different teams represent -- to a good extent -- 
``independent'' CMB maps. We will use this hypothesis in order to estimate residual
foreground contaminations by comparing these different foreground-cleaned maps.

As a matter of fact, the WMAP science team has made substantial efforts to 
improve the data products by minimizing the contaminating effects caused by diffuse 
galactic foregrounds, astrophysical point-sources, artifacts from the instruments and 
measurement process, and systematic errors~\cite{Jarosik:2006ib,Gold:2008kp}. 
As a result, multi-frequency foreground-cleaned full-sky CMB maps were produced, 
named Internal Linear Combination maps, corresponding to three and 
five year WMAP data~\cite{Hinshaw:2006ia,Hinshaw:2008kr}. 
Therefore, in order to account for 
the mentioned randomness, systematic, and contaminating effects of the 
CMB data, we include in our analyses several full-sky foreground-cleaned 
CMB maps, listed in Table~\ref{tab:mapas-ceu-int}, which were produced 
using the three and five year WMAP data.

\begin{table}[h]
\begin{centering}
\begin{tabular}{cc}
\toprule 
Full-sky maps & References\tabularnewline
\midrule
\midrule 
Hinshaw \textit{et. al.} & \cite{Hinshaw:2006ia,Hinshaw:2008kr}\tabularnewline
\midrule 
de Oliveira-Costa \textit{et. al.} & \cite{deOliveiraCosta:2006zj}\tabularnewline
\midrule 
Kim \textit{et. al.} & \cite{Kim:2008zh}\tabularnewline
\midrule 
Park \textit{et. al.}  & \cite{Park:2006dv}\tabularnewline
\midrule 
Delabrouille \textit{et. al.} & \cite{Delabrouille:2008qd}\tabularnewline
\bottomrule
\end{tabular}

\caption{Full-sky foreground cleaned CMB maps from WMAP data used in our analysis 
to estimate $a_{\ell m}^{(\scriptsize{\mbox{obs}})}$.
Note that the reference \cite{Kim:2008zh} includes the analysis 
of maps from the three and five years WMAP releases.}
\label{tab:mapas-ceu-int}
\par\end{centering}
\end{table}

The prescription we adopt to determine the distribution of the observational
variable $y$ can be equally applied either to 
partial or full-sky CMB maps, and is achieved as follows: we simulate Gaussian
random $a_{\ell m}$'s in such a way that {\it their central values are
given by the five year ILC5 data} \cite{Hinshaw:2006ia,Hinshaw:2008kr},
and with a variance which is estimated from the \textit{sample standard
deviation} of all the maps listed in Table~\ref{tab:mapas-ceu-int}. 
So, for example, suppose we have $n$ different full-sky temperature maps at hand and
we want to estimate the randomness inherent in the determination of,
let's say, $a_{32}$. Therefore, we take:
\begin{equation}
\mathcal{N}(a_{32}^{\scriptsize{\mbox{ILC5}}},\sigma_{32})\;\rightarrow\; a_{32} \, ,
\label{protocolo-1}
\end{equation}
with:
\begin{equation}
\sigma_{32}=\sqrt{\frac{1}{n-1}\sum_{i=1}^{n}(a_{32}^{i}-\bar{a}_{32})^{2}}\qquad\mbox{and}
\qquad\bar{a}_{32}=\frac{1}{n}\sum_{i=1}^{n}a_{32}^{i} \, ,
\label{protocolo-2}
\end{equation}
where $\mathcal{N}(\mu,\sigma)$ represents a Gaussian distribution
with mean $\mu$ and standard deviation $\sigma$.

Our procedure may be justified by the following facts: first, the ILC5 data is 
undoubtly the most accurate CMB measurement we currently have at our disposal, and 
for this reason it can be considered as a good approximation to the ``real'' temperature 
map. Secondly, as we already mentioned, by comparing different 
CMB maps we should have an estimate of the residual contamination
which were not properly removed, and that can possibly be the source 
of anomalies. Note that if this contamination is indeed weak, then the sample 
variance above will be small, and our procedure will reduce to the standard way
of calculating probabilities (see the discussion of $\S$\ref{sec:Probabilities}).
Finally, the use of a Gaussian in (\ref{protocolo-1}) was dictated not only by 
simplicity, but also by the fact that the propagation of uncertainties in 
physical experiments is usually assumed to follow a normal distribution.

\subsection{Full-sky maps}
\begin{figure}[h]
\begin{center}
\includegraphics[scale=0.4]{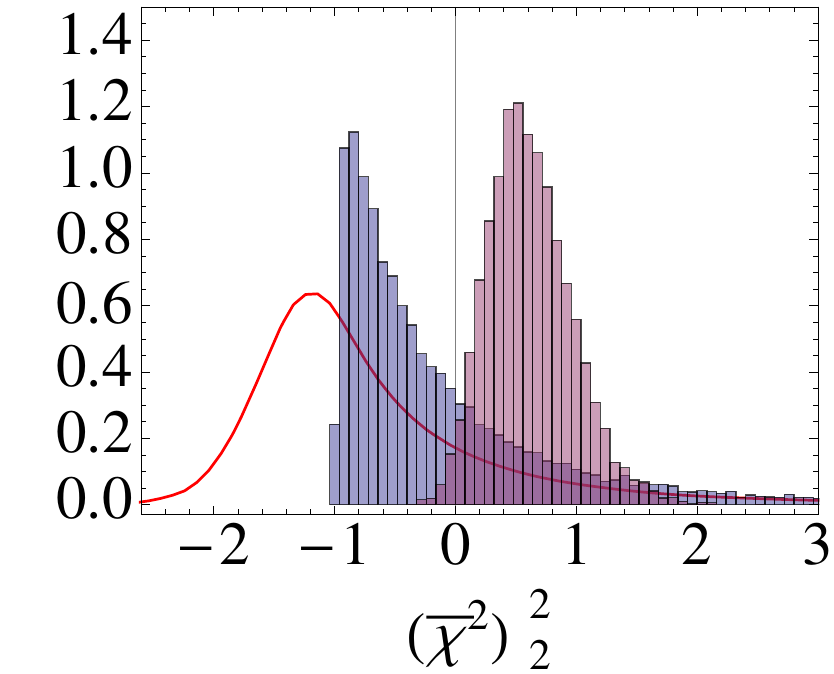} \includegraphics[scale=0.4]{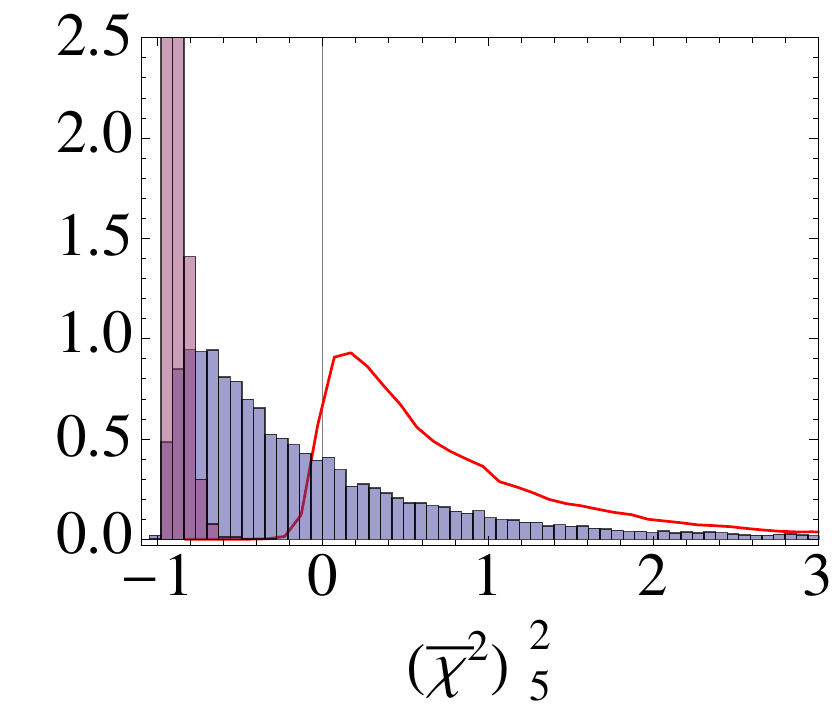}
\includegraphics[scale=0.4]{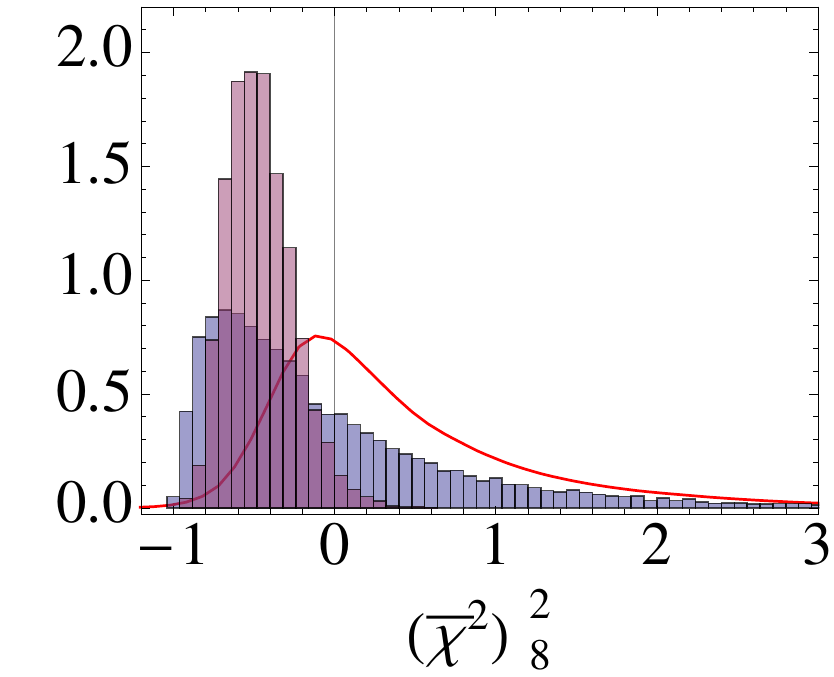} \includegraphics[scale=0.4]{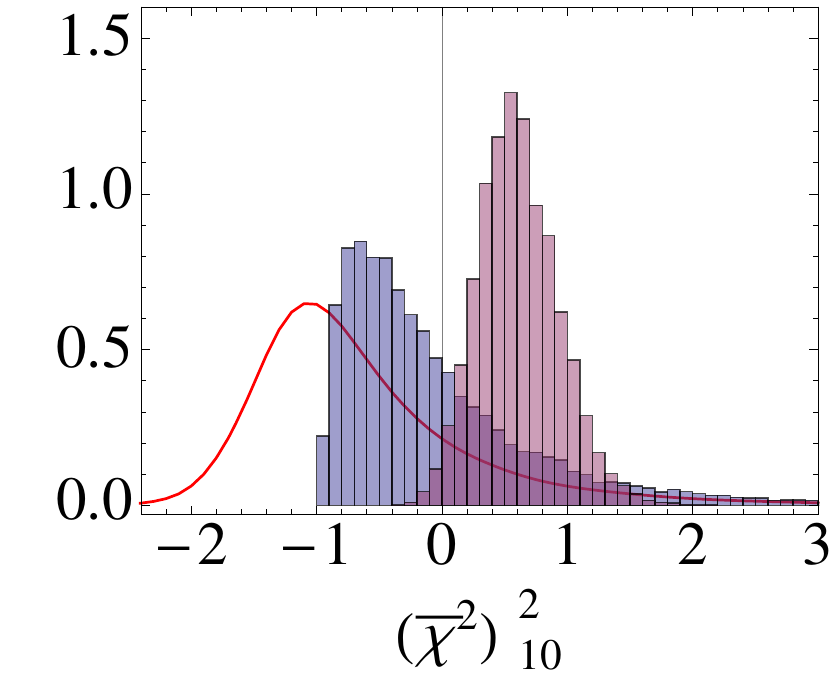}

\includegraphics[scale=0.4]{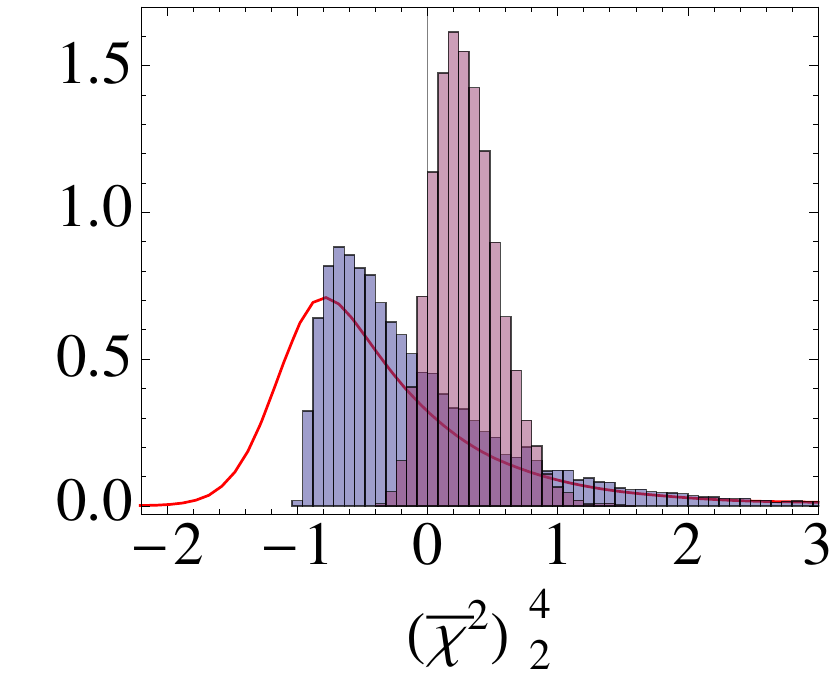} \includegraphics[scale=0.4]{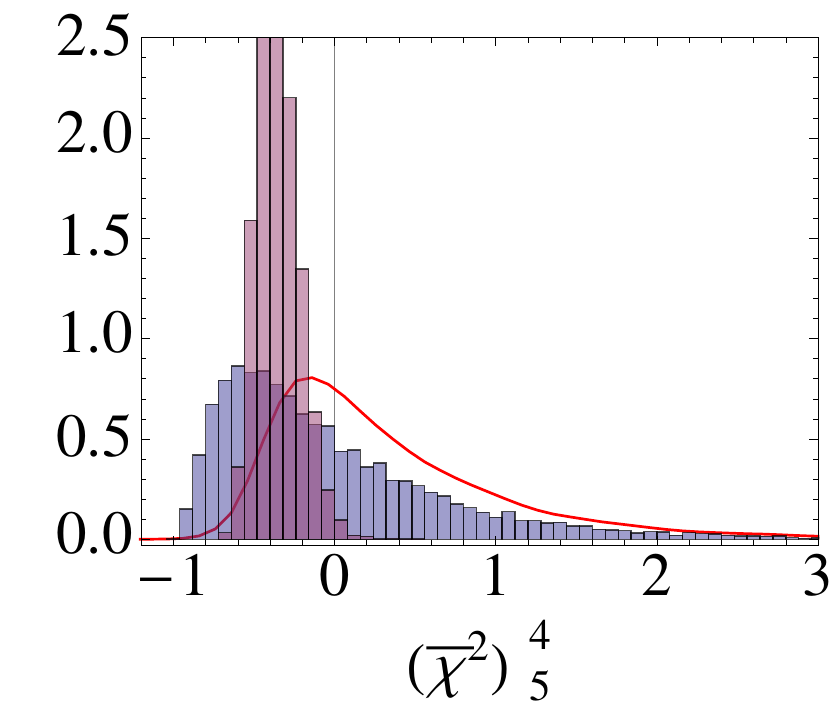}
\includegraphics[scale=0.4]{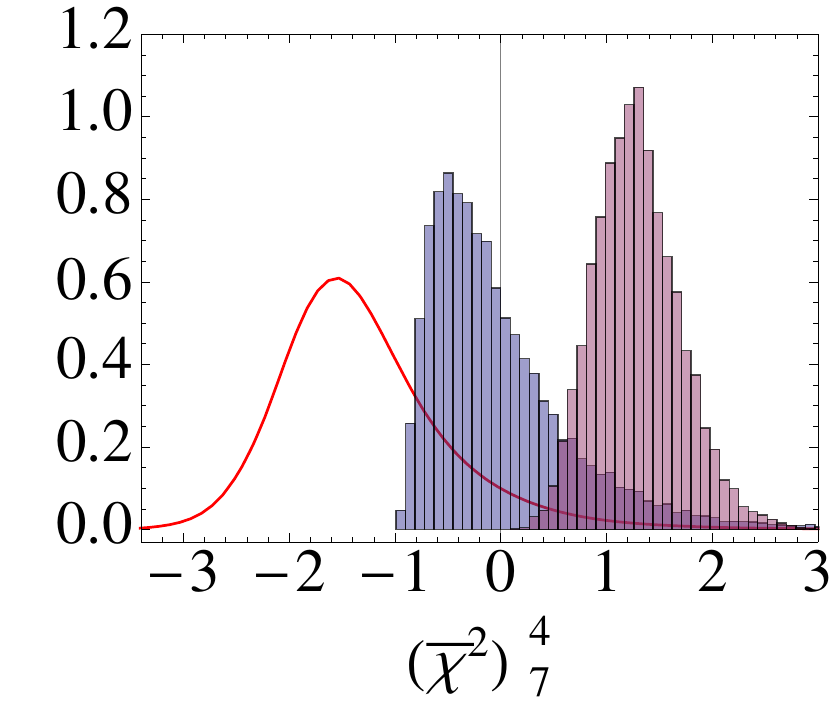} \includegraphics[scale=0.4]{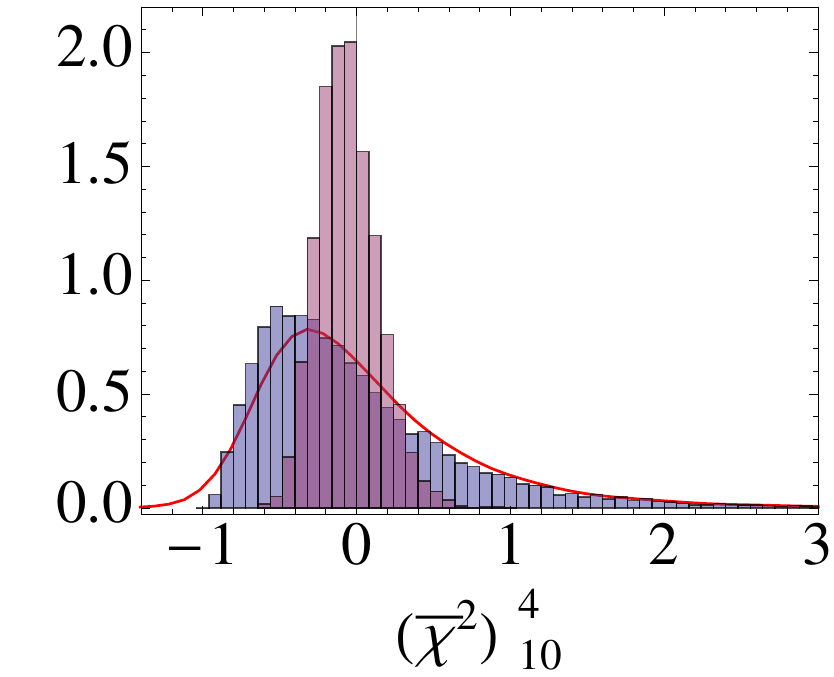}

\includegraphics[scale=0.4]{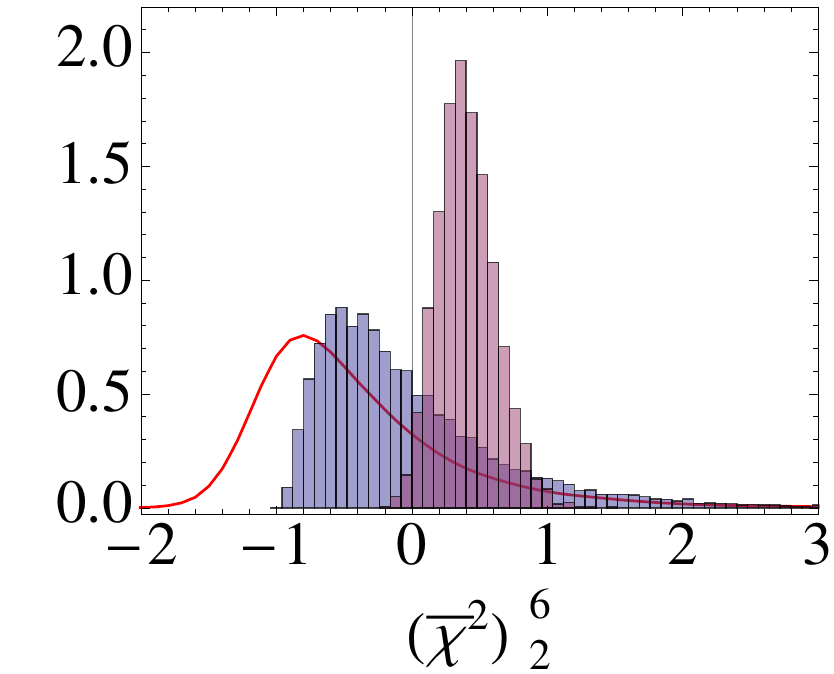} \includegraphics[scale=0.4]{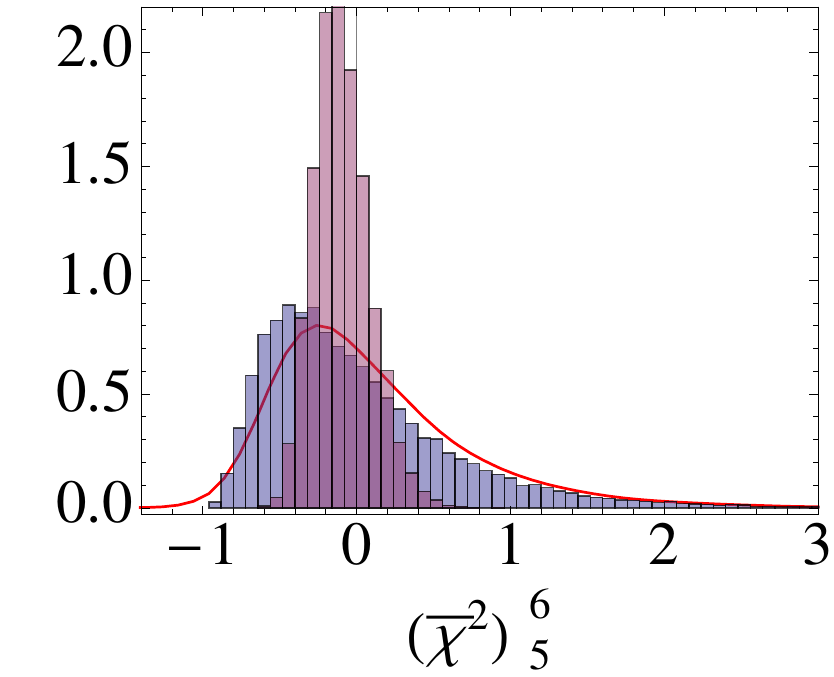}
\includegraphics[scale=0.4]{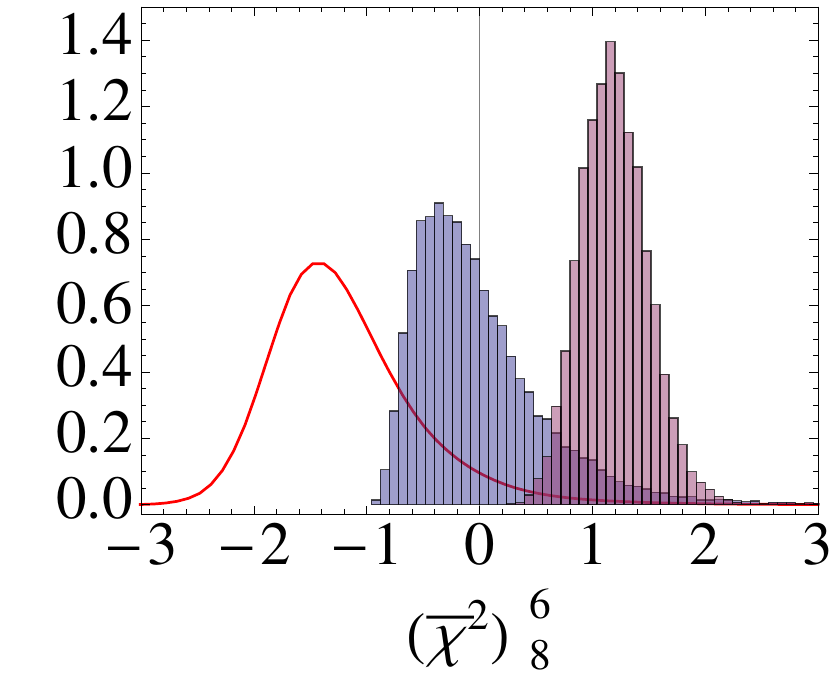} \includegraphics[scale=0.4]{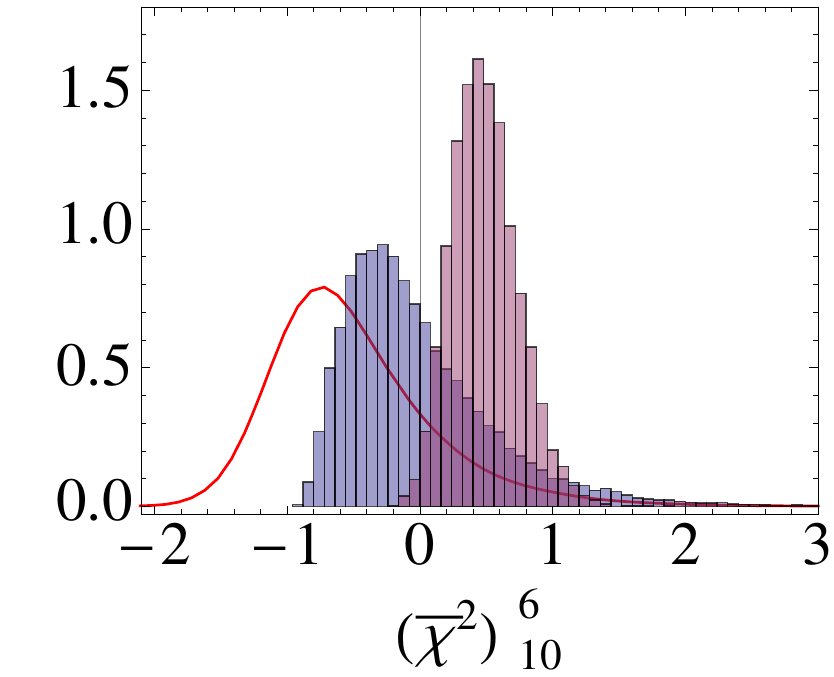}

\includegraphics[scale=0.4]{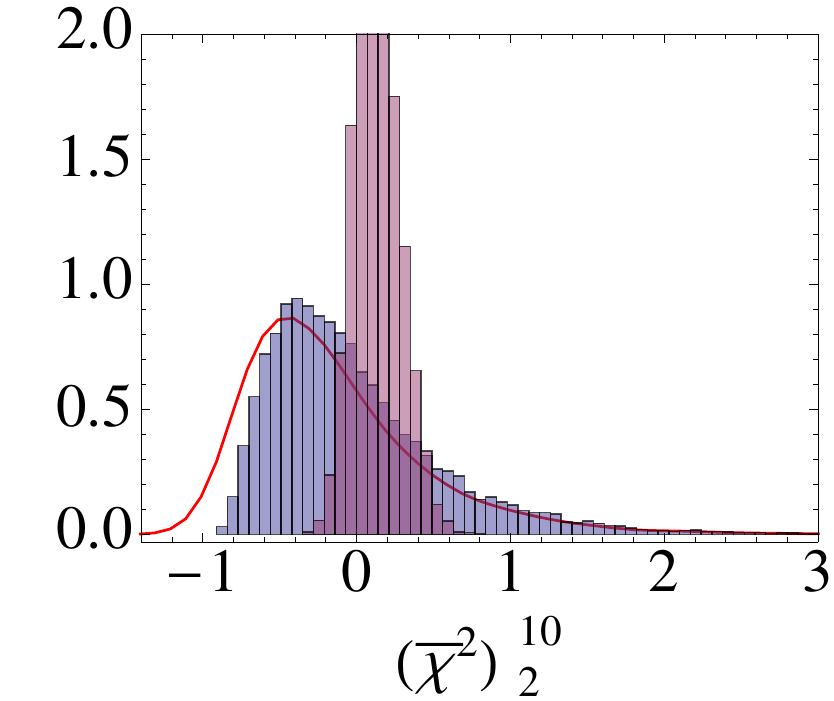} \includegraphics[scale=0.4]{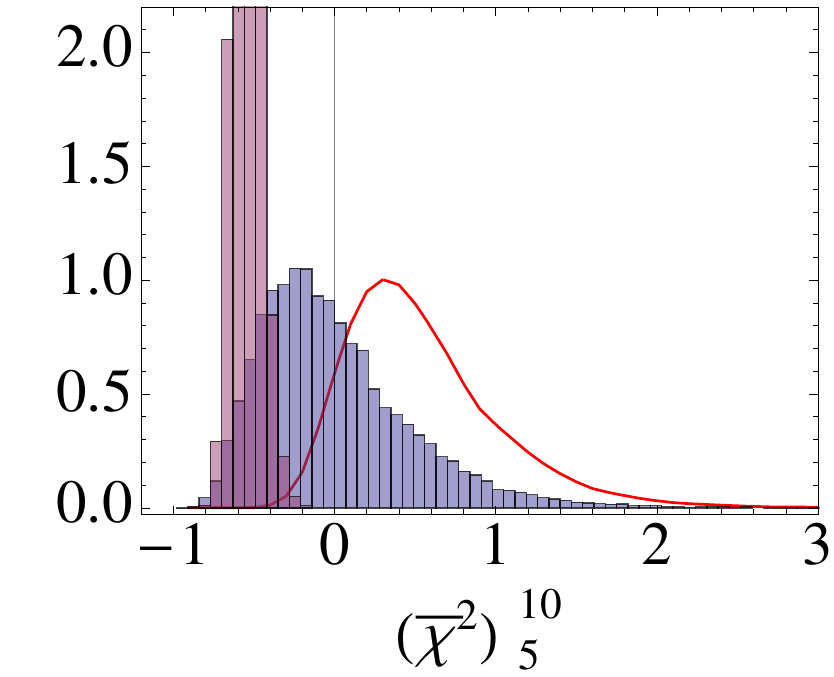}
\includegraphics[scale=0.4]{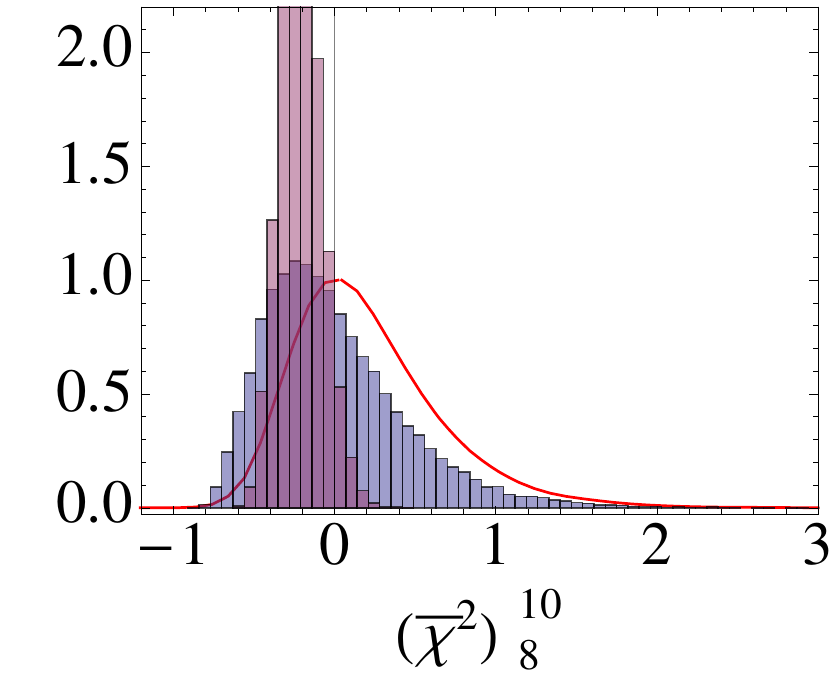} \includegraphics[scale=0.4]{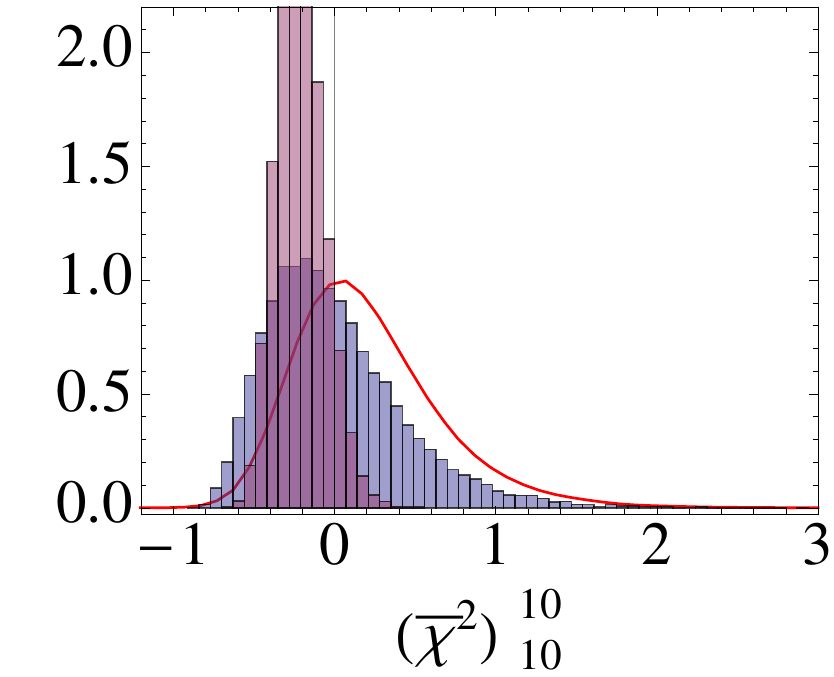}

\caption{Full-sky maps pdf's for $(\bar{\chi}^2)_{\ell(\scriptsize{\mbox{th}})}^{l}$
(blue histograms), $(\bar{\chi}^2)_{\ell(\scriptsize{\mbox{obs}})}^{l}$
(purple histograms) and for the difference 
$(\bar{\chi}^2)_{\ell(\scriptsize{\mbox{th}})}^{l} - (\bar{\chi}^2)_{\ell(\scriptsize{\mbox{obs}})}^{l}$
(solid red line). We show only a few representative figures since the
remaining ones are qualitatively the same. The final probabilities are shown in Table 
(\ref{tab:prob-finais-ceuint}) and correspond to the area under the solid curve from 
$-\infty$ to $0$. All pdf's are normalized to 1.\label{fig:hist-ceuint}}
\end{center}
\end{figure}

Following this procedure, we have used the full-sky maps shown in
Table (\ref{tab:mapas-ceu-int}) to construct $10^{4}$ Gaussian random
$a_{\ell m}$'s, which were then used to calculate $10^{4}$ realizations of 
$y=(\bar{\chi}^2)^l_{\ell({\rm obs})}$. With those variables we constructed histograms
which, together with the histograms for the (full-sky) variable $x$, were used to calculate
the final probability (\ref{P>}). We have restricted our analysis
to the range of values $(\ell,l)\in[2,10]$, since the low multipolar sector 
(i.e., large angular scales) is where most of the anomalies were reported. 
The resulting histograms and pdf's are shown in Fig. (\ref{fig:hist-ceuint}), and 
the final probabilities we obtained are show in Table (\ref{tab:prob-finais-ceuint}).

\begin{table}[H]
\begin{centering}
\begin{tabular}{cccccccccc}
\toprule 
$l\backslash\ell$ & \multicolumn{1}{c}{2} & \multicolumn{1}{c}{3} & 4 & 5 & 6 & 7 & 8 & 9 & 10\tabularnewline
\midrule
\midrule 
2 & 81.1\% & 73.7\% & 54.1\% & \textbf{6.1}\% & 80.7\% & 46.4\% & 36.9\% & 47.5\% & 81.8\%\tabularnewline
\midrule 
4 & 74.0\% & 72.6\% & 55.0\% & 39.6\% & 74.2\% & \textbf{93.2\%} & 51.6\% & 55.4\% & 56.3\%\tabularnewline
\midrule 
6 & 78.1\% & 80.7\% & 69.3\% & 52.3\% & 33.6\% & 80.0\% & \textbf{95.0}\% & 50.3\% & 82.2\%\tabularnewline
\midrule 
8 & 63.5\% & 87.7\% & 18.8\% & 51.5\% & 21.4\% & 66.4\% & 31.6\% & 27.5\% & 82.3\%\tabularnewline
\midrule 
10 & 67.9\% & 50.0\% & 61.0\% & \textbf{8.7}\% & 37.7\% & 59.5\% & 36.6\% & 29.2\% & 35.7\%\tabularnewline
\bottomrule
\end{tabular}

\caption{Final probabilities of obtaining, in a random $\Lambda$CDM universe, a chi-square 
value smaller or equal to $(\bar{\chi}^2)_{\ell(\scriptsize{\rm obs})}^{l}$, as given by 
full-sky temperature maps.
\label{tab:prob-finais-ceuint}}\par
\end{centering}
\end{table}

It is interesting to compare the values in Table (\ref{tab:prob-finais-ceuint})
with the more naive and independent `error-bars' estimate made
in~\cite{Pereira:2009kg}. As anticipated there, the angular quadrupole $\ell=2$
does not seem anomalous under the tests we considered. The only curiosity
lies in the fact that, for the scales we probed, the quantity $(\bar{\chi}^2)_{2}^{l}$ 
has a probability which is always greater than 50\%, meaning that the quadrupole $\ell=2$ 
has a consistently positive planar modulation. 

By far, the most unlikely individual values in Table (\ref{tab:prob-finais-ceuint})
are in the sectors $(l,\ell)$ given by $(2,5)$, $(10,5)$, $(4,7)$
and $(6,8)$, and are all below a relative chance of 10\% of either
being too negative [$\left(2,5\right)$, $\left(10,5\right)$] or too positive
[$(4,7)$, $(6,8)$]. Although these modulations do not extend to
a significant range of $l$ values, it is possible that these sectors
are contaminated by foreground residuals. If this is the case, then
these anomalies should not be present in masked temperature maps. 
Conversely, if these are the only anomalies present in CMB data, then the 
next cut-sky maps analyses should not reveal other extreme probability 
values (i.e., lower than 10\% or higher than 90\%). 
We will analyze these questions in more detail in the next Section.

But before we turn to the masked maps, it is important to realize that
not only the individual values of $(\bar{\chi}^2)^l_\ell$ are relevant:
their coherence over a range of angular or planar momenta also 
carries interesting information.
So, for example, a set of $(\bar{\chi}^2)^l_\ell$'s which are all individually within
the cosmic variance bounds, but which are all positive (or negative)
can be an indication of an excess (or lack) of modulation. In fact, as 
we have shown in \S\ref{sec:exemplo}, if one suppresses a Gaussian 
and isotropic map only over a thick disk between fixed latitudes, the effect is to
boost/depress the modulations compared with the isotropic case,
over a large range of $l$'s and $\ell$'s. The $l$'s and
$\ell$'s which are most affected are the ones corresponding to the angular
scales subtended by the disk, but all momenta $l$ and $\ell$ of
the angular-planar spectra are affected, with the net effect that they
gain a positive/negative bias.

This type of coherent behavior (an apparent bias) is in fact exactly what we observe in 
the following cases: $(\bar{\chi}^2)^l_2$, $(\bar{\chi}^2)^l_3$ and, to a lesser extent, 
$(\bar{\chi}^2)^4_\ell$
-- see Table \ref{tab:prob-finais-ceuint}. The angular quadrupole $\ell=2$,
as well as the angular octopole $\ell=3$, have all positive planar spectra (for all
values of $l$ which we were able to compute), indicated by probabilities larger
than 50\%.
The planar hexadecupole $l=4$ also has 8 out of 9 angular spectra assuming
positive values (only $\ell=5$ is negative). 
Again, we stress that the data analyzed in this Section relates
to the full-sky maps, which are certainly still affected by residual galactic foregrounds, 
so it is important that we only draw conclusions after looking into the same results for 
the cut-sky (masked) maps -- the subject of our next Section.

\subsection{Masked maps\label{sec:masks}}

\begin{figure}[H]
\begin{centering}
\includegraphics[scale=0.4]{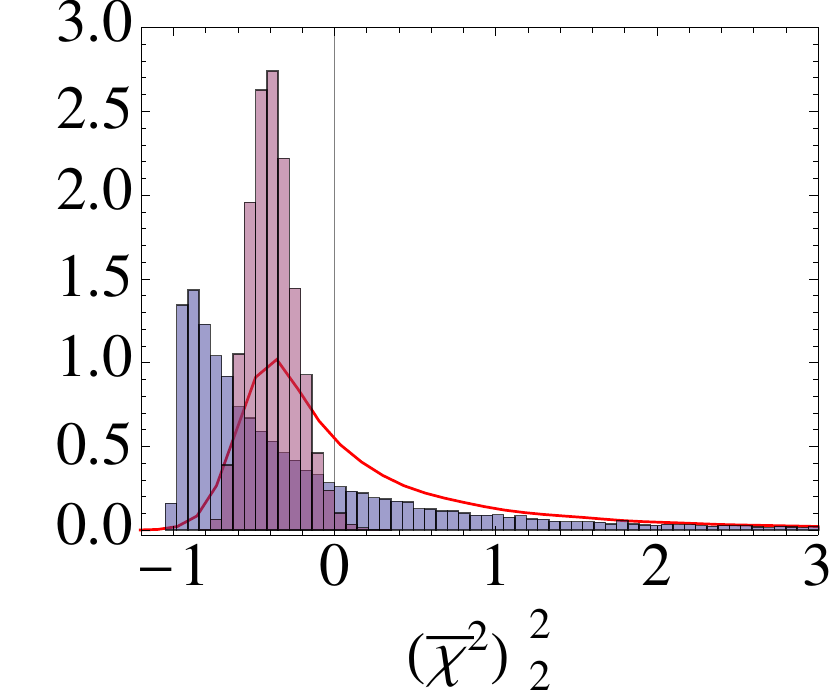} \includegraphics[scale=0.4]{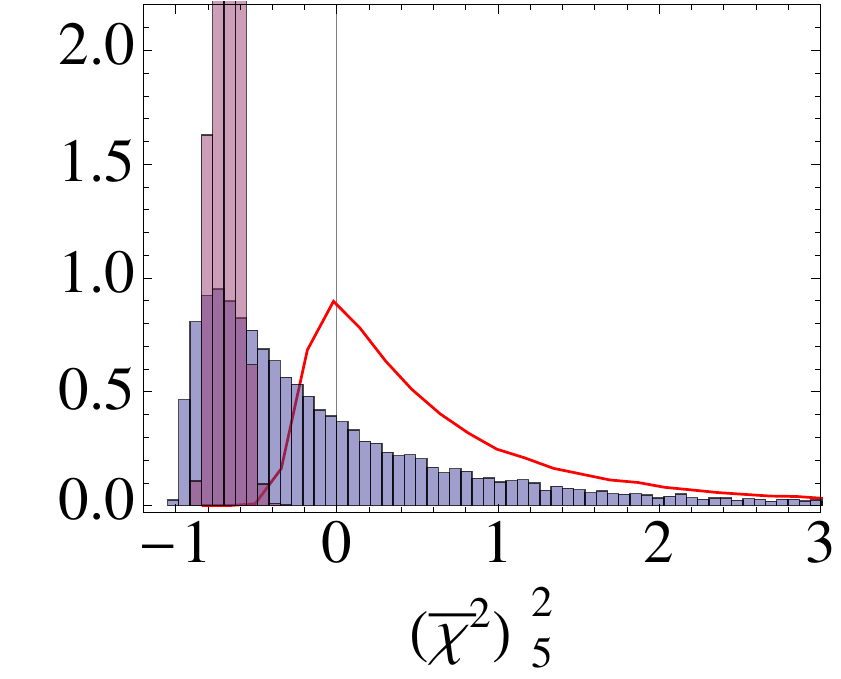}
\includegraphics[scale=0.4]{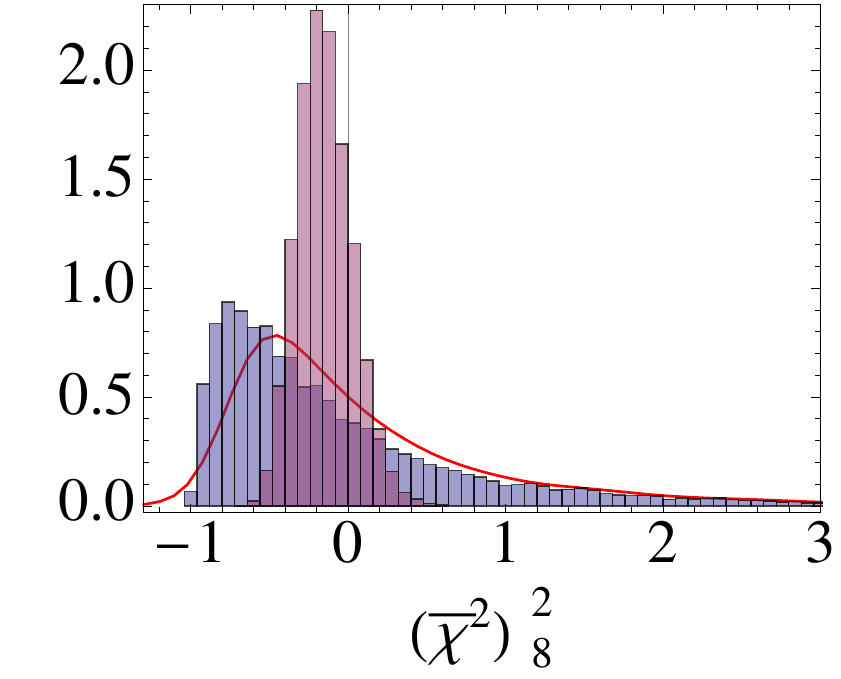} \includegraphics[scale=0.4]{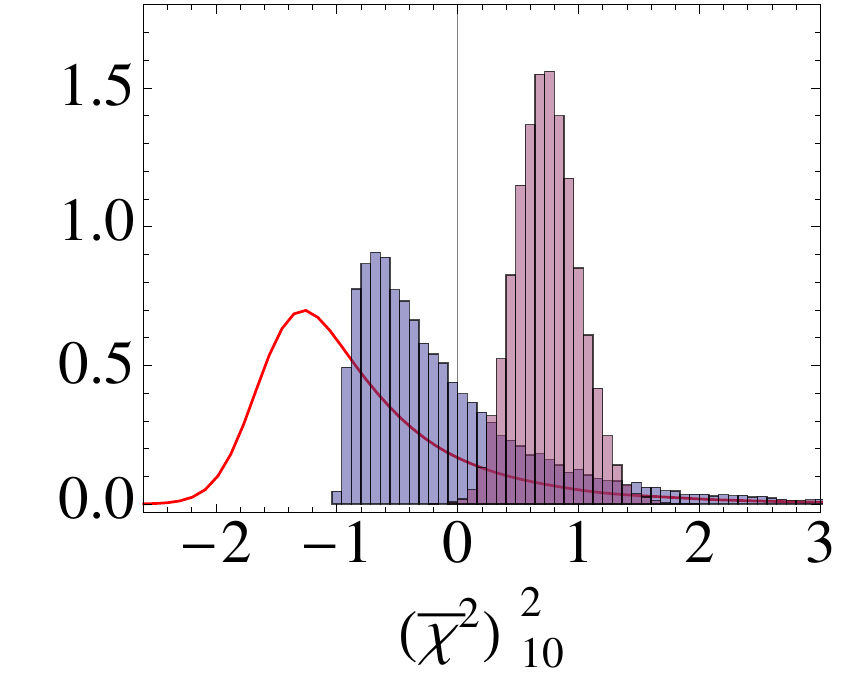}

\includegraphics[scale=0.4]{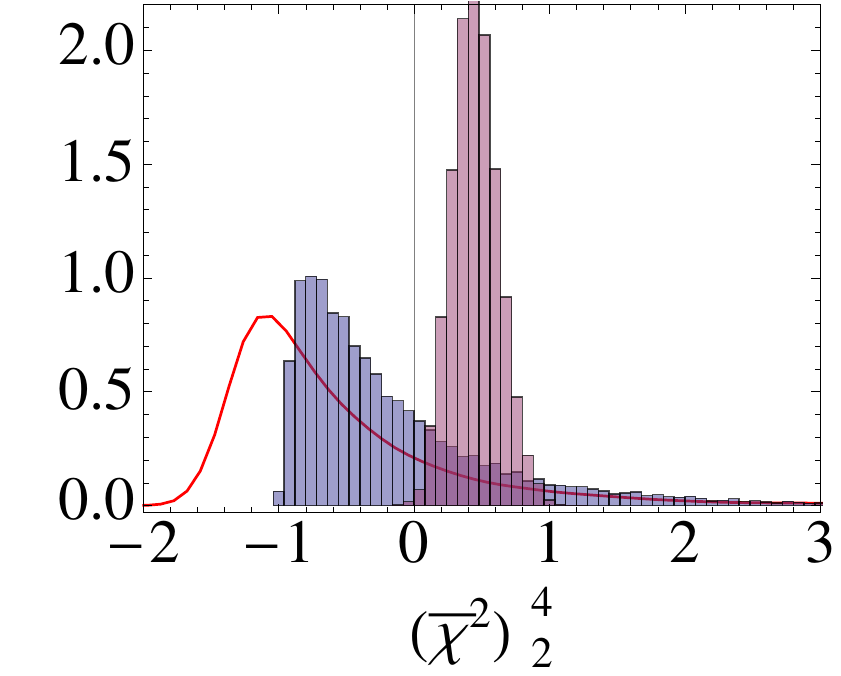} \includegraphics[scale=0.4]{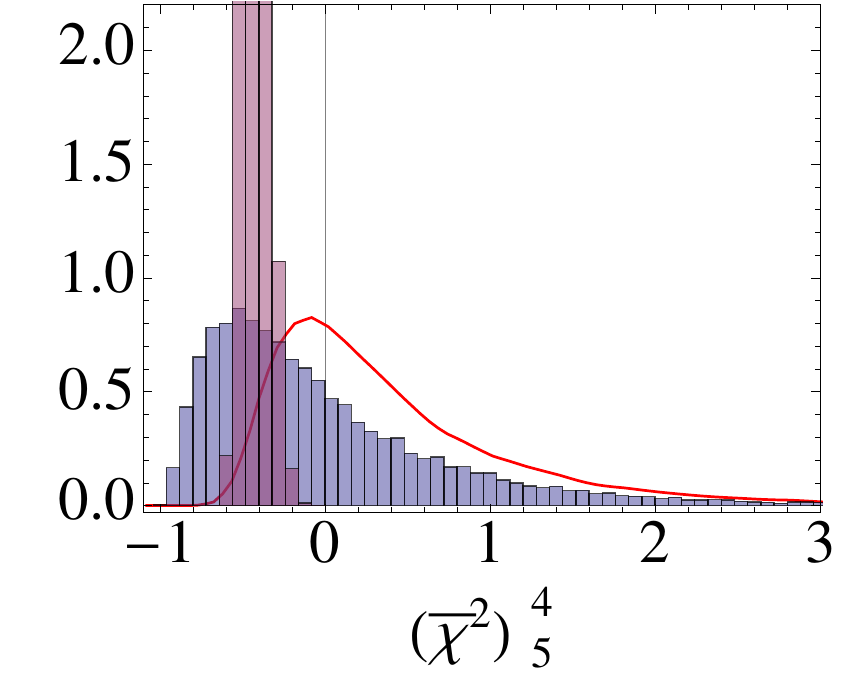}
\includegraphics[scale=0.4]{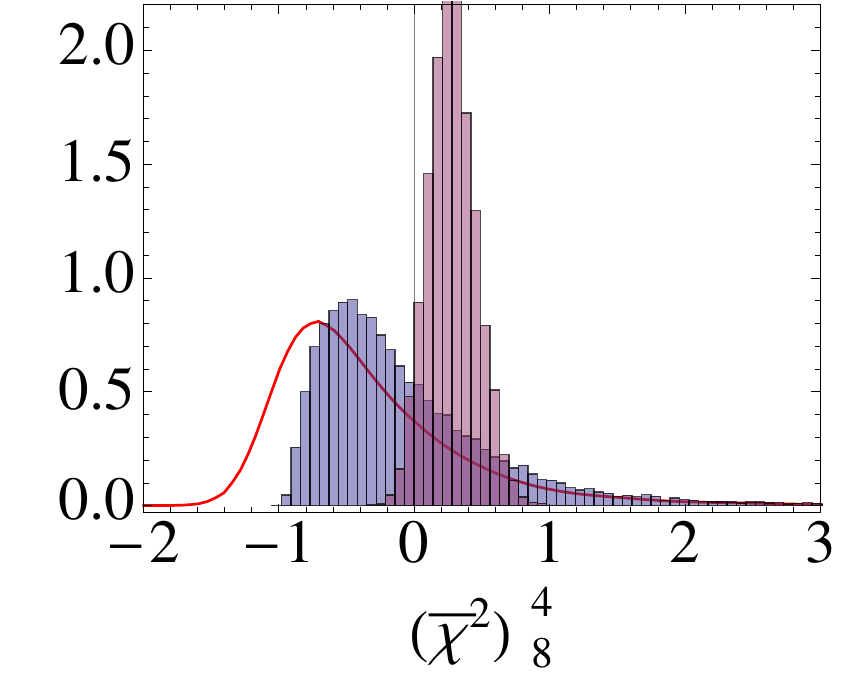} \includegraphics[scale=0.4]{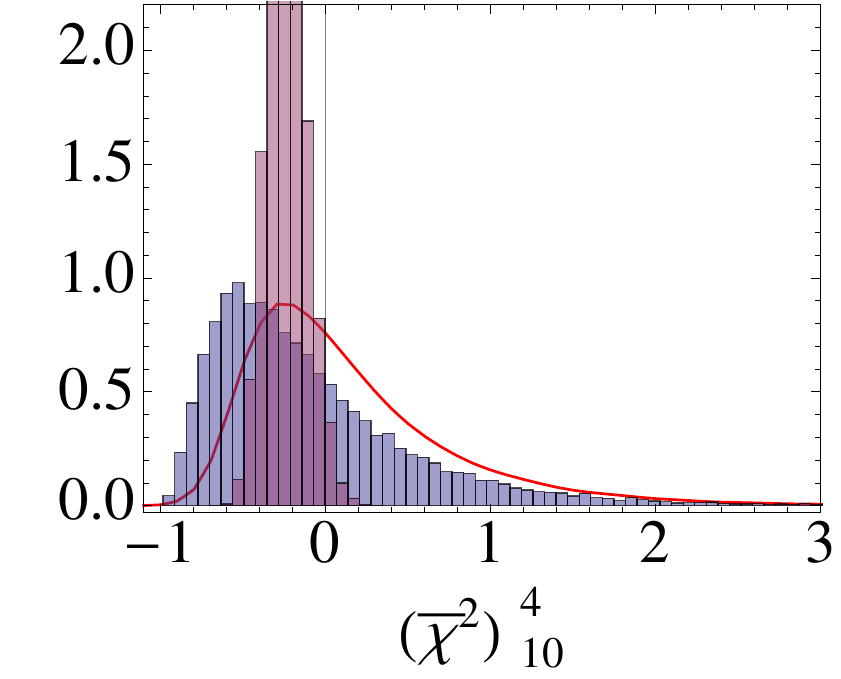}

\includegraphics[scale=0.4]{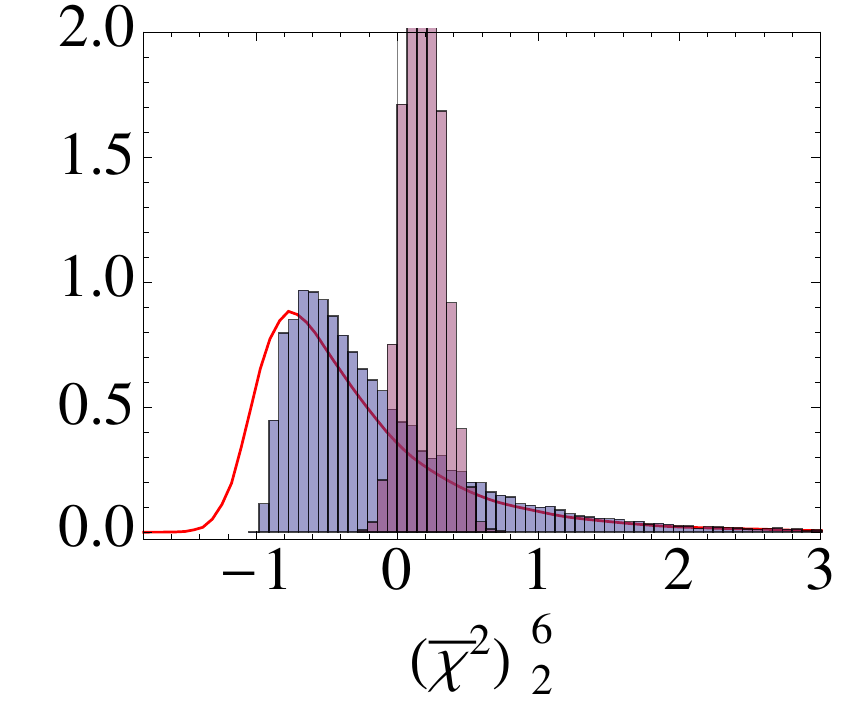} \includegraphics[scale=0.4]{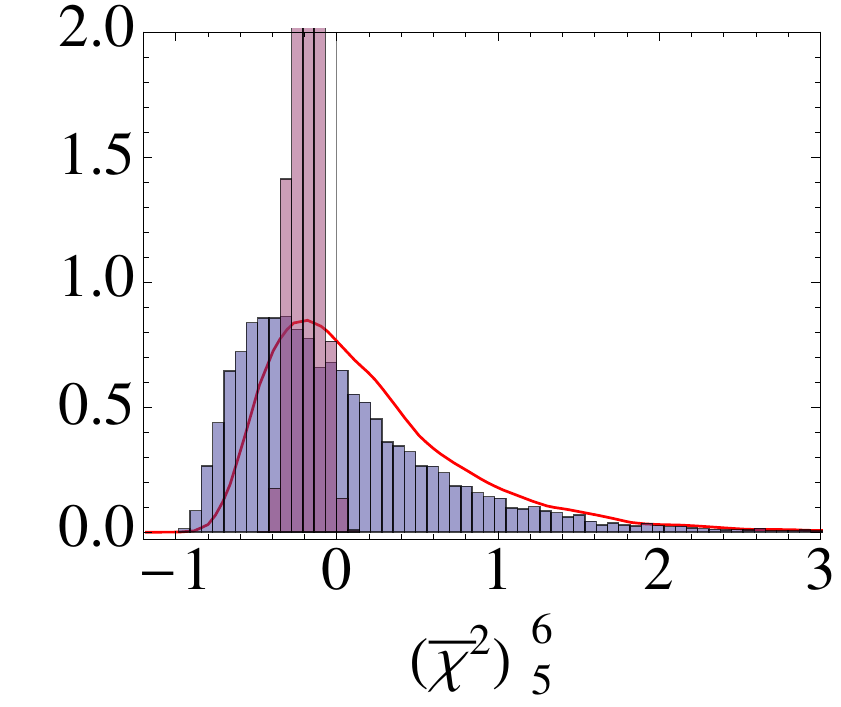}
\includegraphics[scale=0.4]{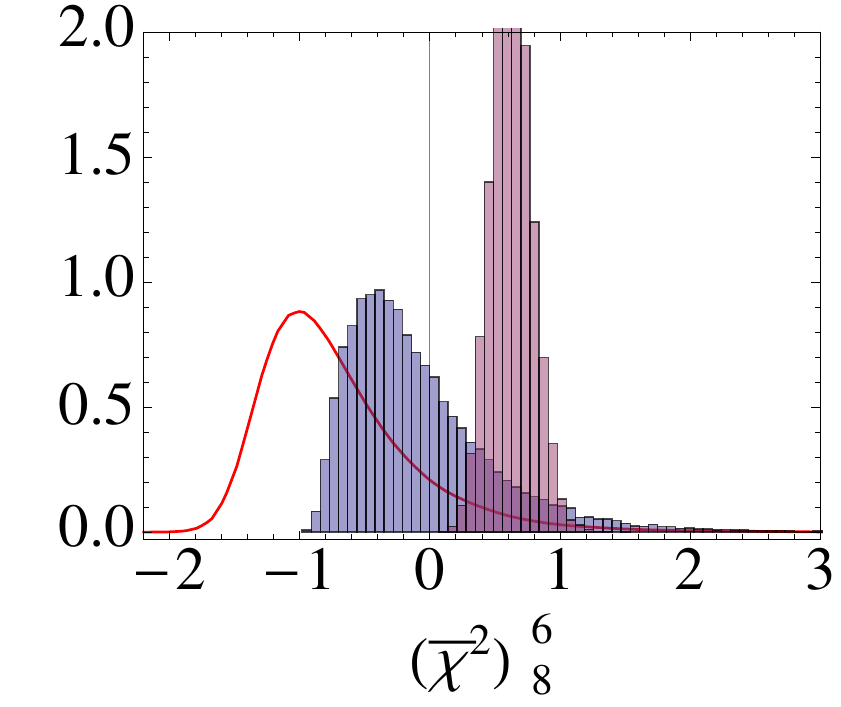} \includegraphics[scale=0.4]{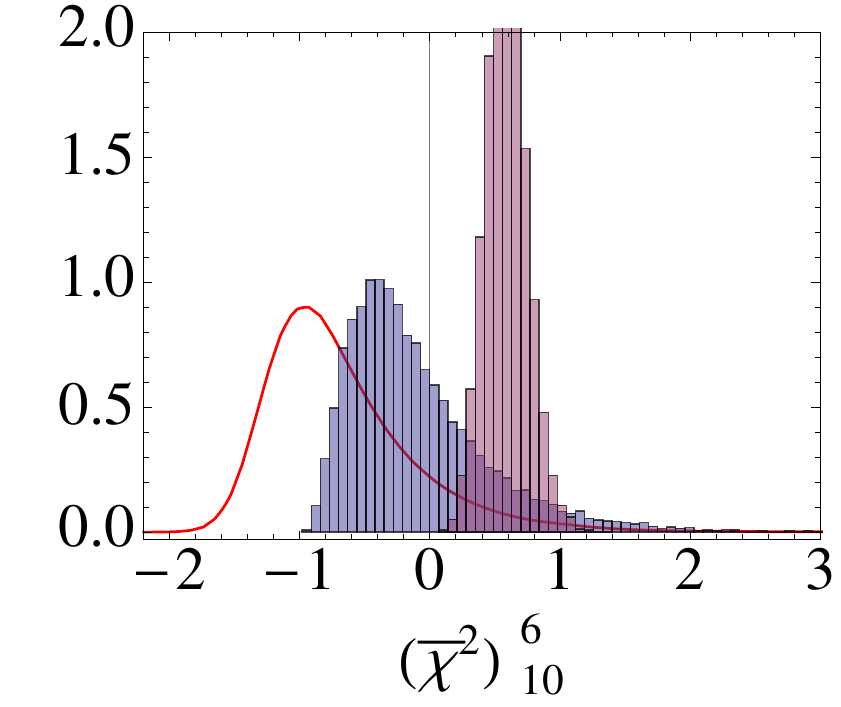}

\includegraphics[scale=0.4]{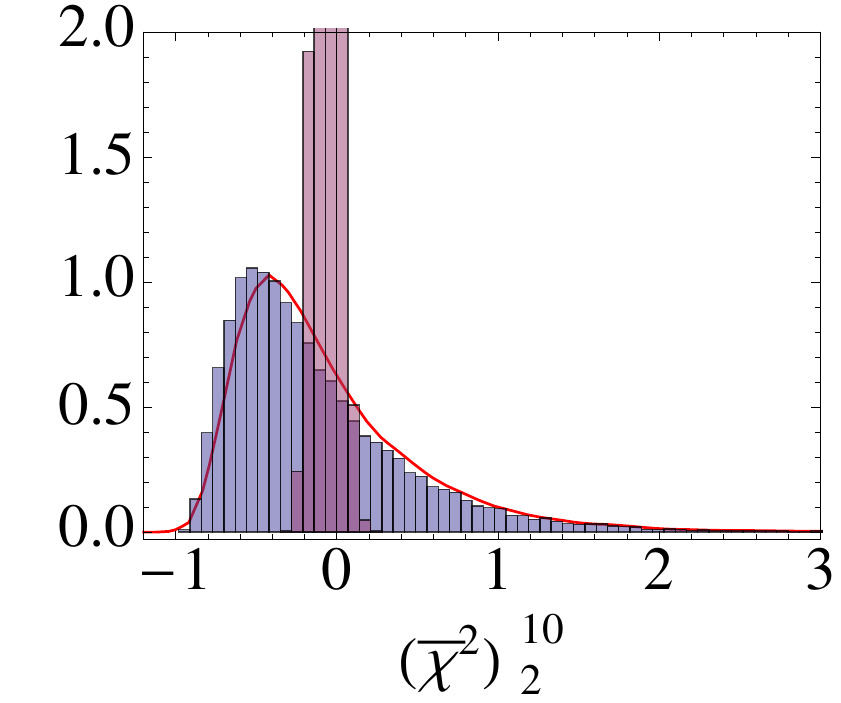} \includegraphics[scale=0.4]{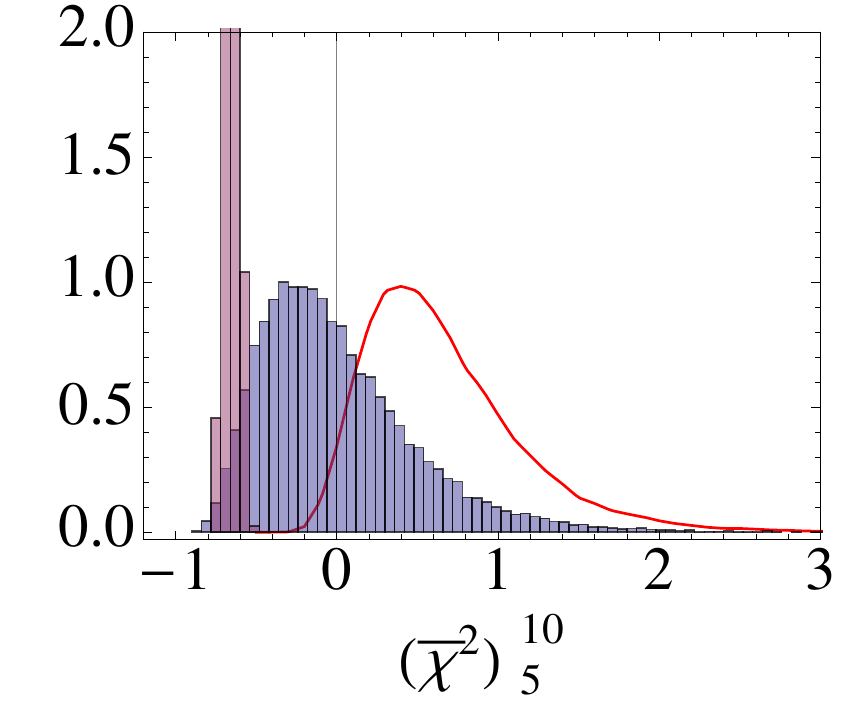}
\includegraphics[scale=0.4]{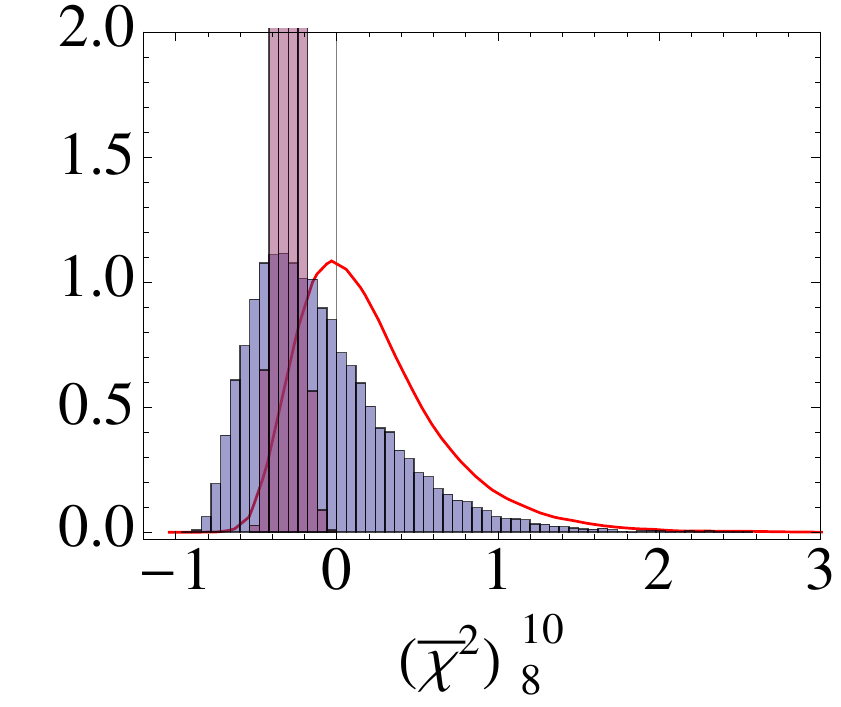} \includegraphics[scale=0.4]{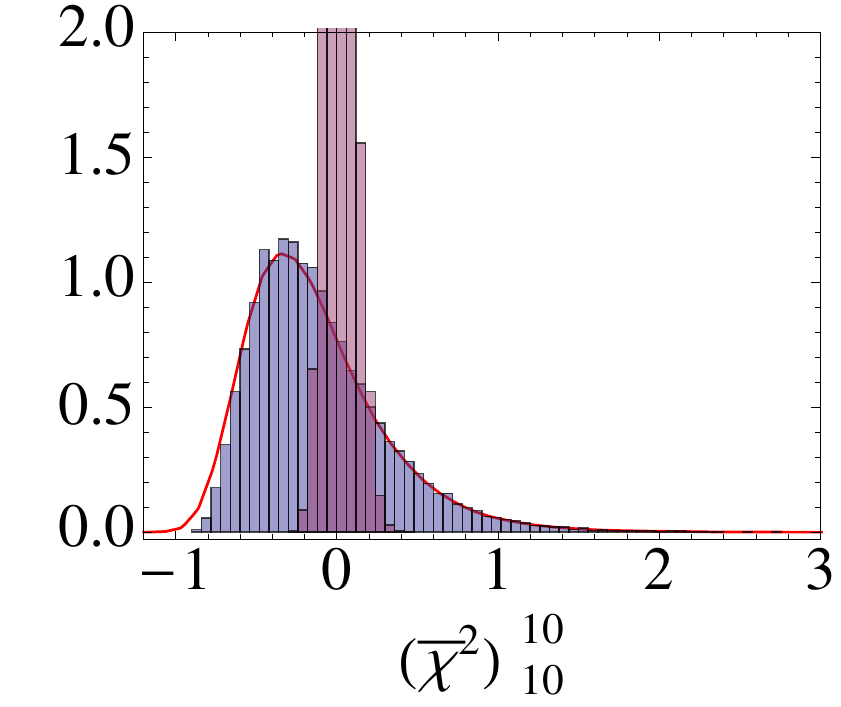}

\end{centering}
\caption{Pdf's for $(\bar{\chi}^2)_{\ell(\scriptsize{\mbox{th}})}^{l}$ (blue histograms),
$(\bar{\chi}^2)_{\ell(\scriptsize{\mbox{obs}})}^{l}$ (purple histograms)
and for the difference 
$(\bar{\chi}^2)_{\ell(\scriptsize{\mbox{th}})}^{l}-
(\bar{\chi}^2)_{\ell(\scriptsize{\mbox{obs}})}^{l}$
(solid red line) for maps with mask KQ85. Again, we only show some representative figures
for simplicity. The final probabilities are shown in Table (\ref{tab:prob-finais-kq85}).
\label{fig:hist-kq85}}
\end{figure}

\begin{figure}[H]
\begin{centering}
\includegraphics[scale=0.4]{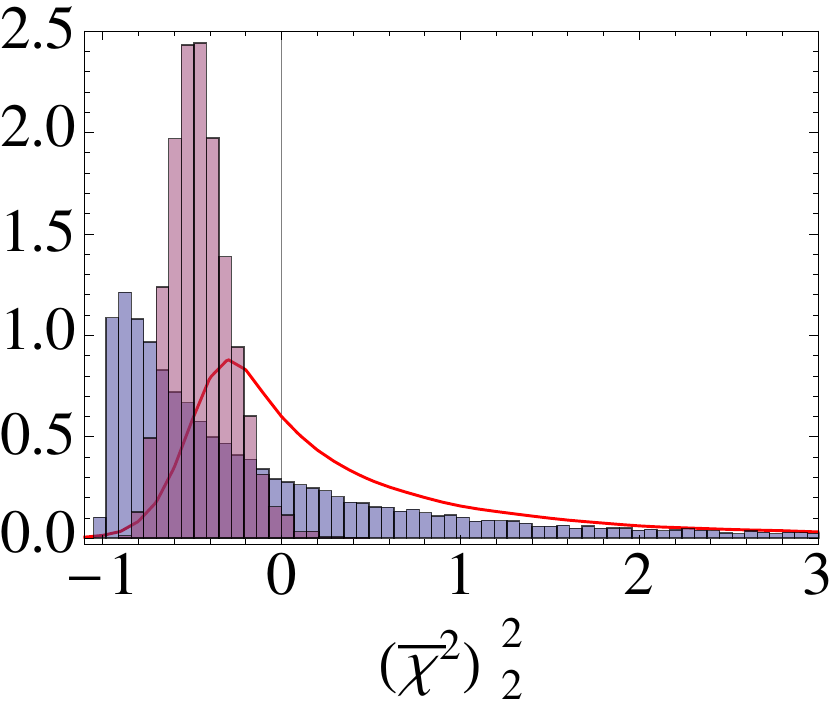} \includegraphics[scale=0.4]{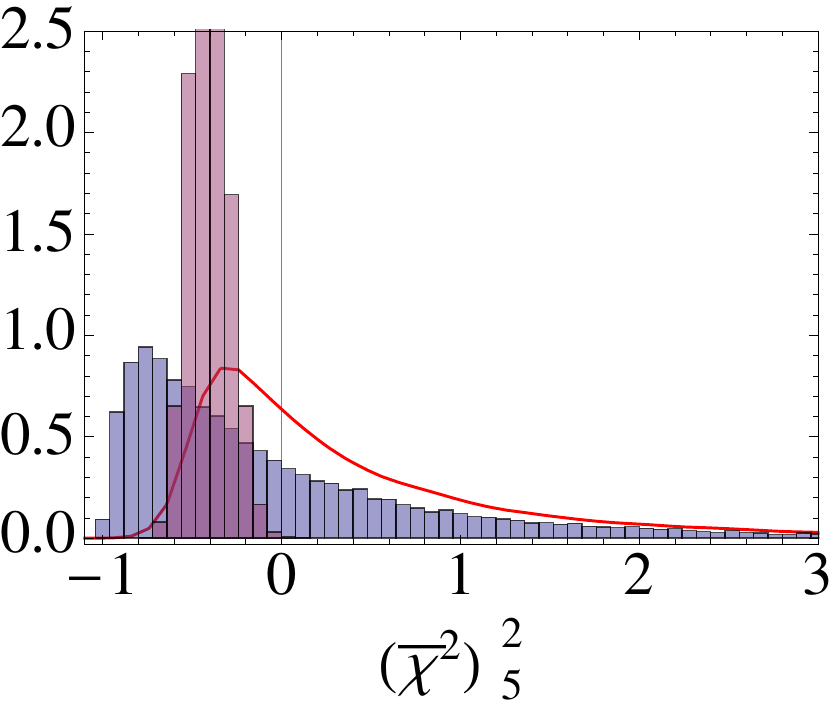}
\includegraphics[scale=0.4]{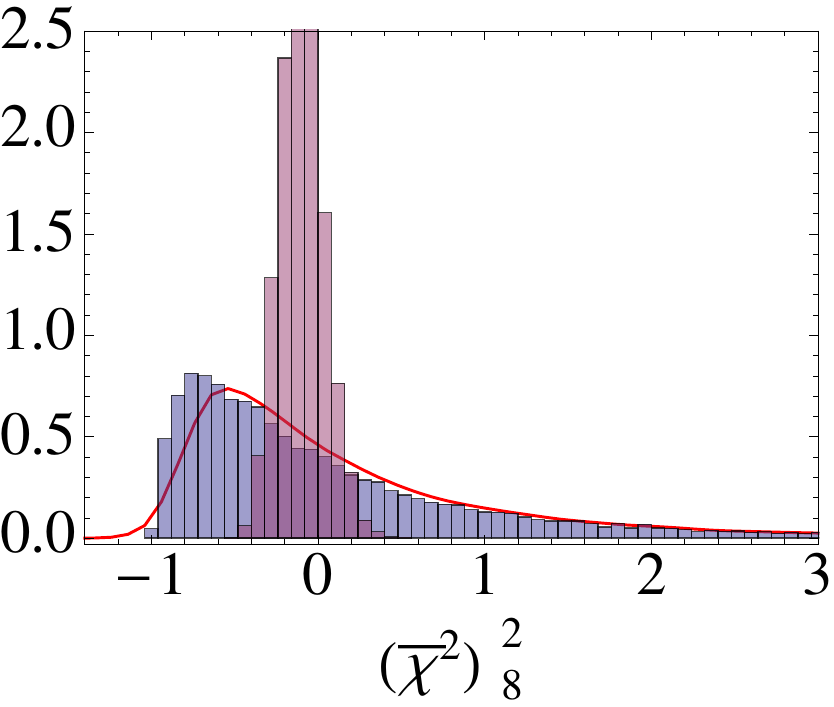} \includegraphics[scale=0.4]{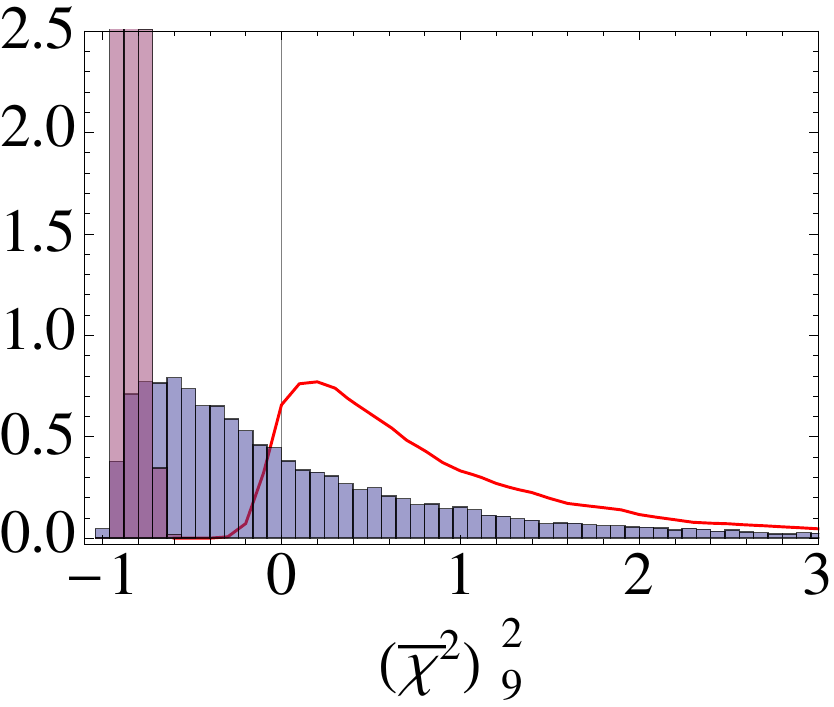}

\includegraphics[scale=0.4]{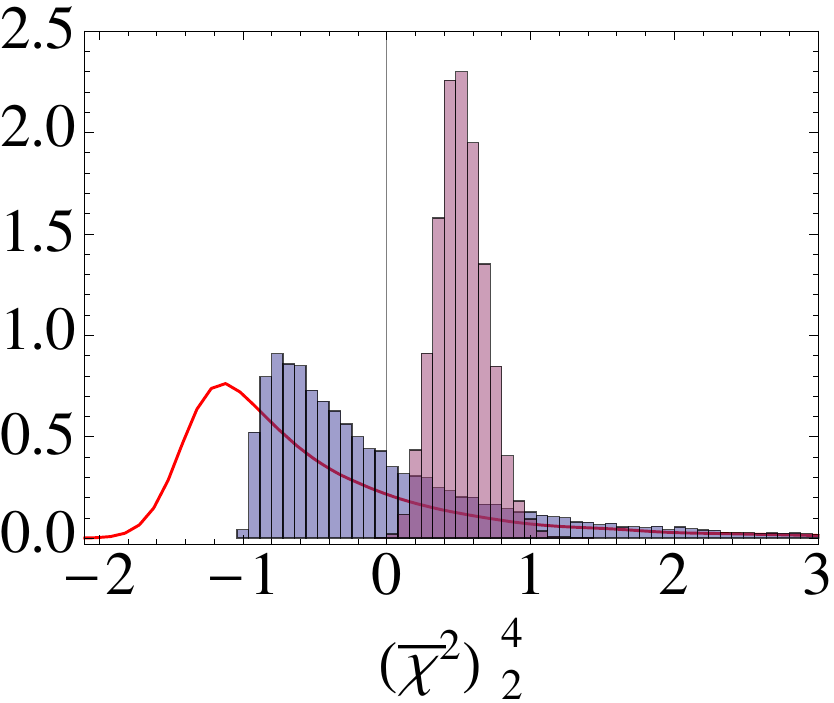} \includegraphics[scale=0.4]{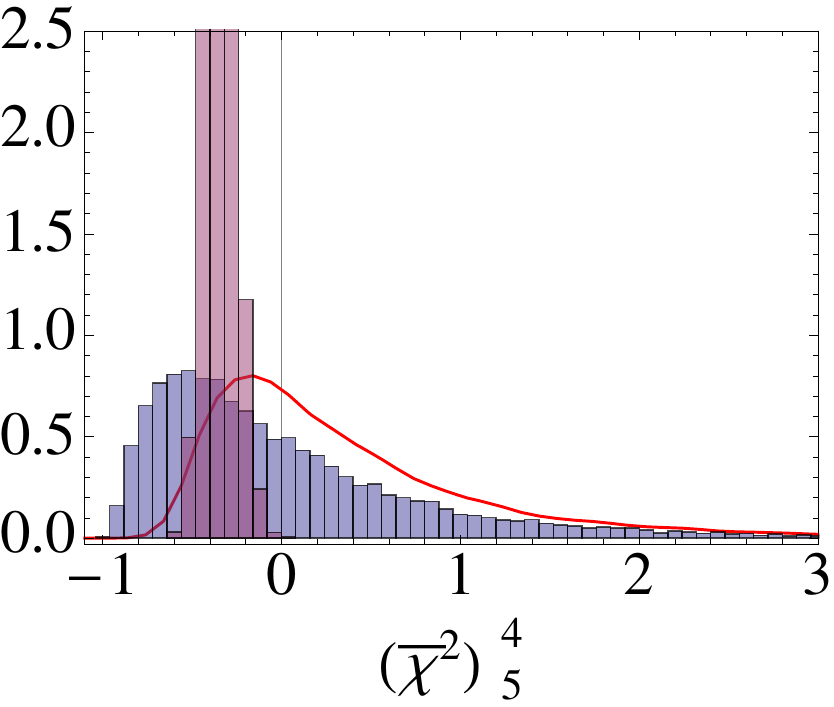}
\includegraphics[scale=0.4]{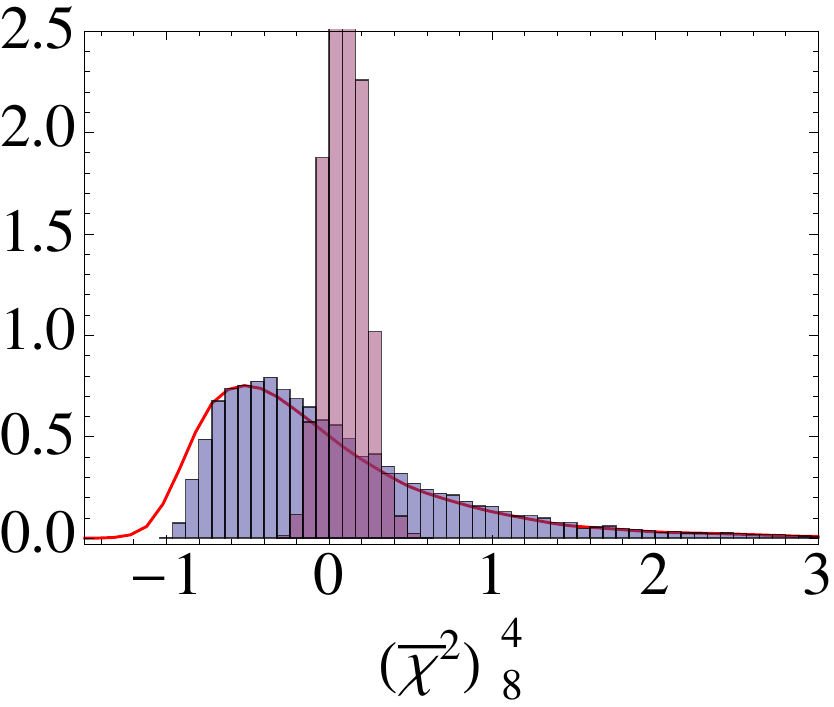} \includegraphics[scale=0.4]{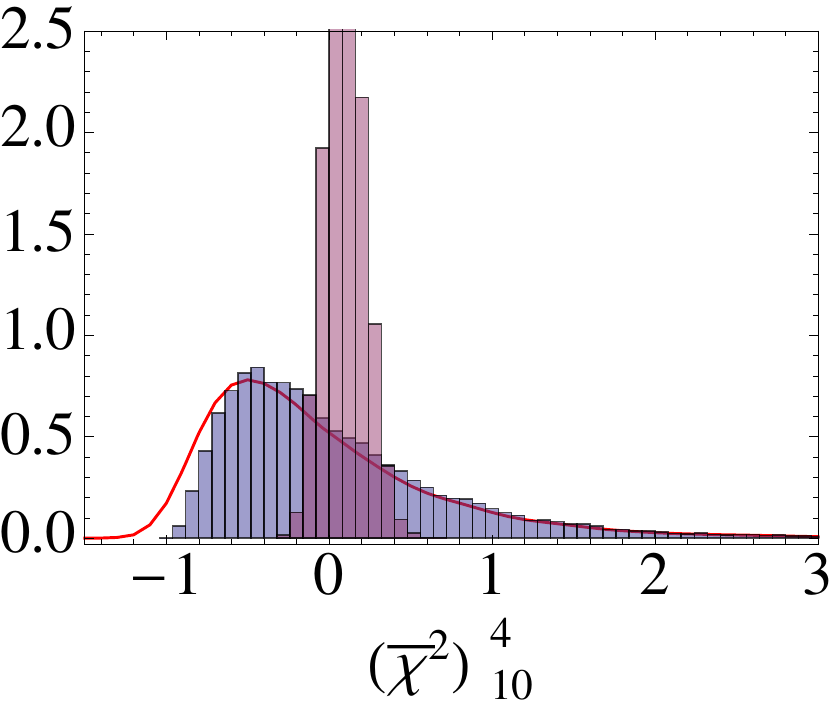}

\includegraphics[scale=0.4]{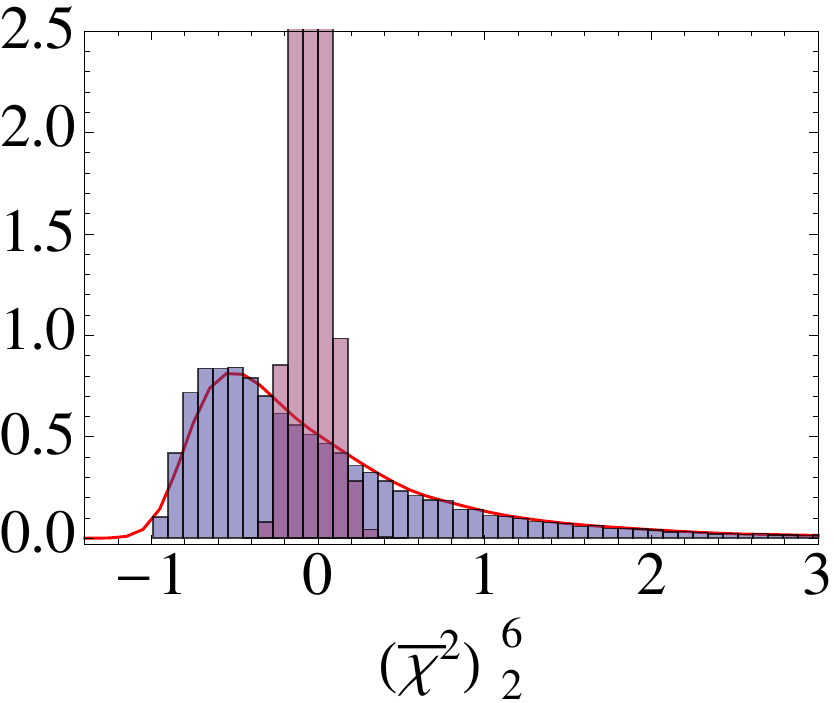} \includegraphics[scale=0.4]{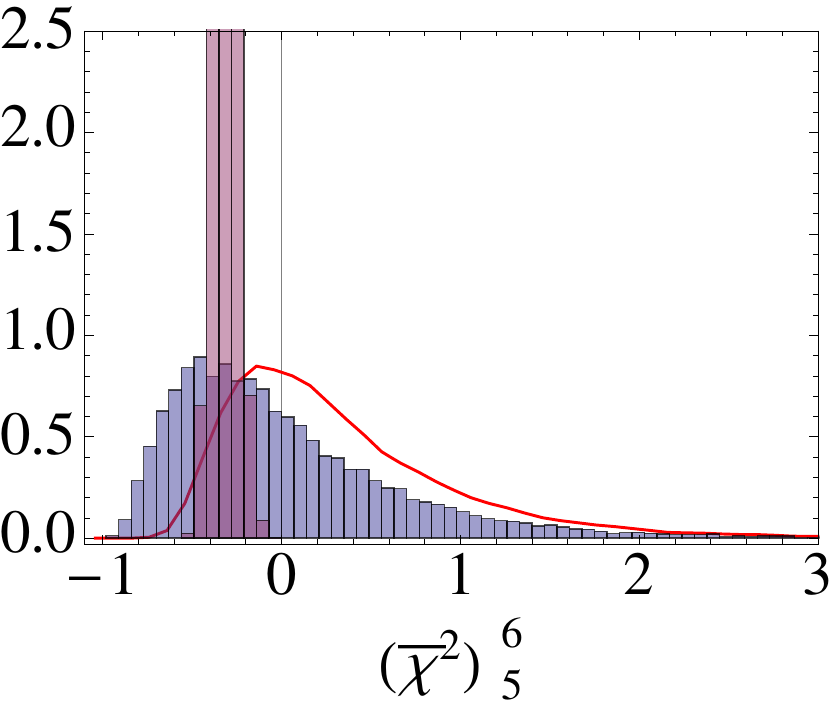}
\includegraphics[scale=0.4]{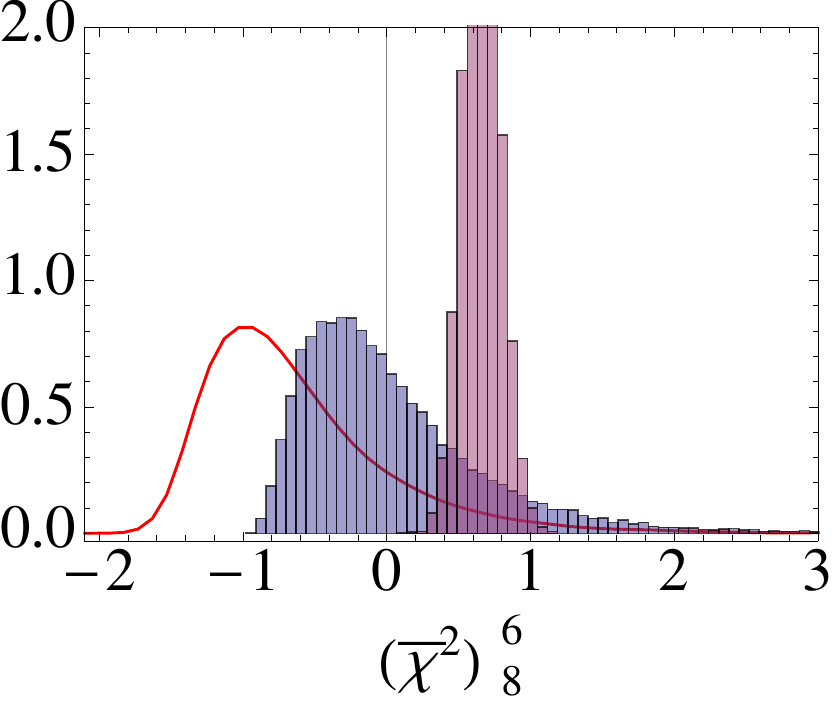} \includegraphics[scale=0.4]{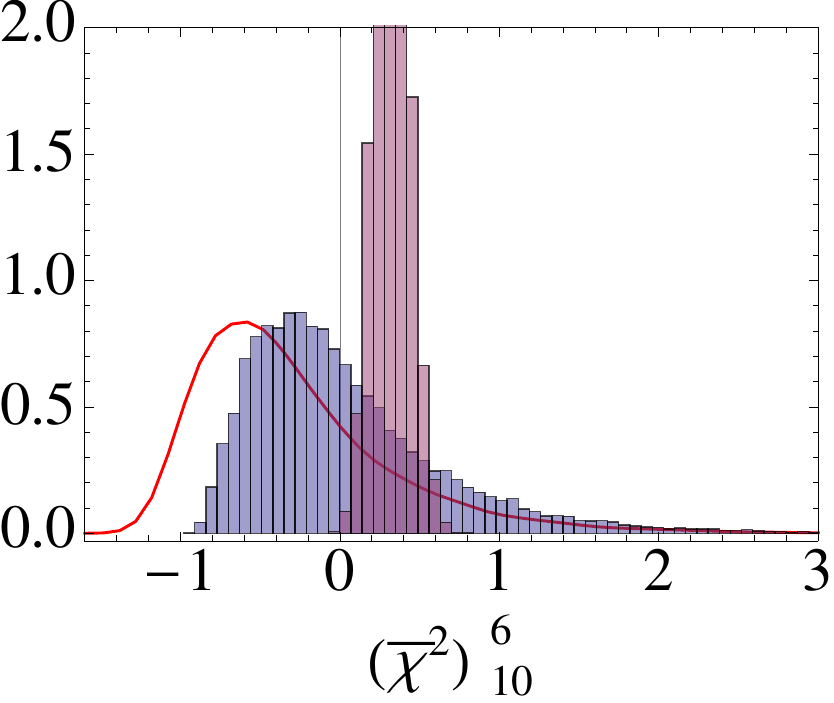}

\includegraphics[scale=0.4]{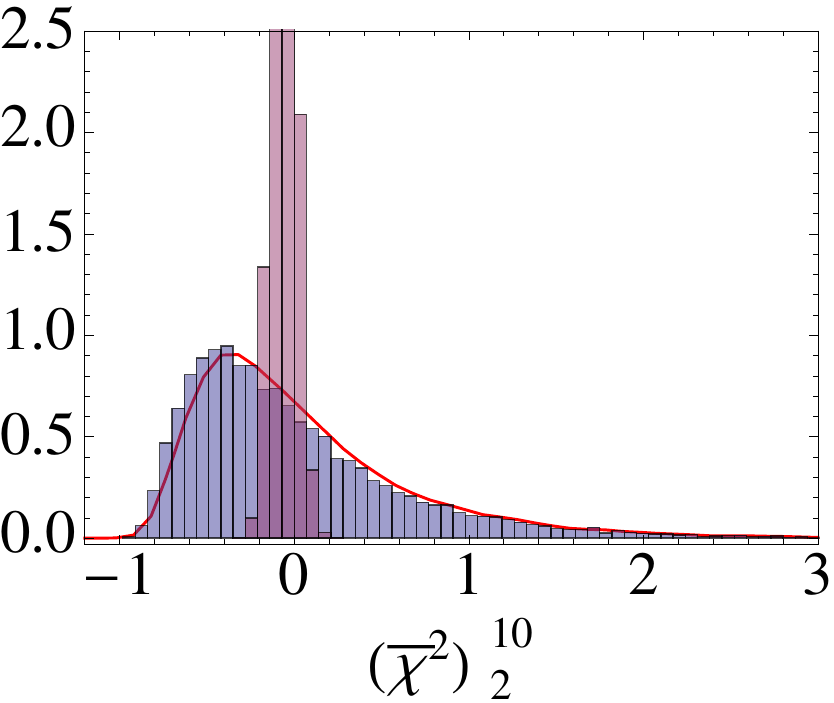} \includegraphics[scale=0.4]{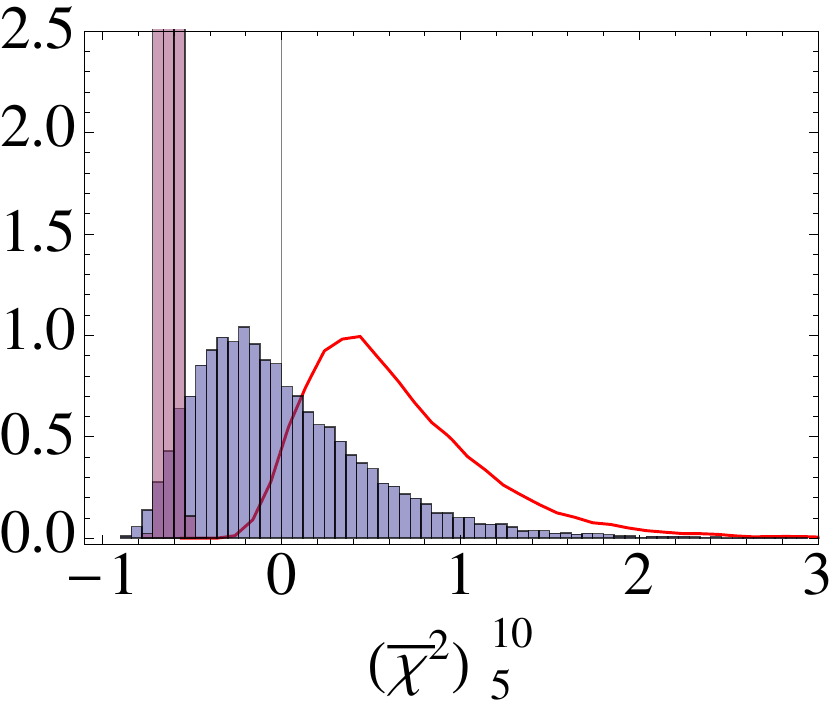}
\includegraphics[scale=0.4]{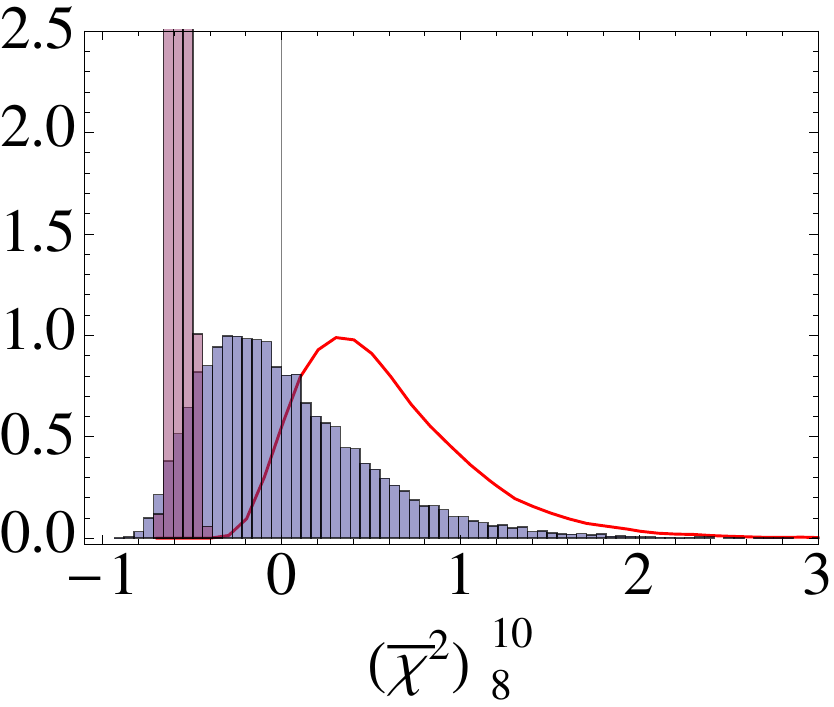} \includegraphics[scale=0.4]{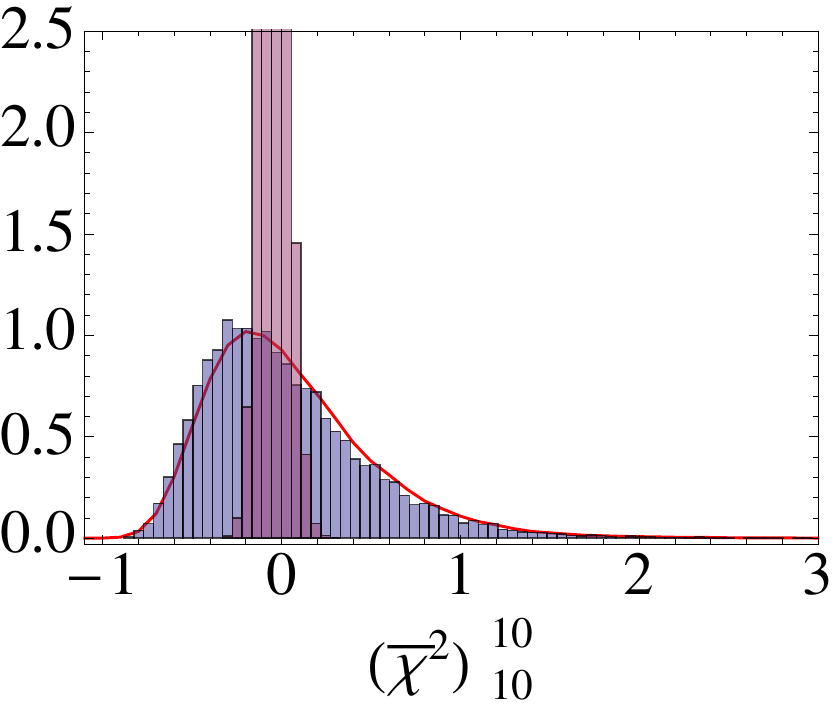}
\par\end{centering}

\caption{Pdf's for $(\bar{\chi}^2)_{\ell(\scriptsize{\mbox{th}})}^{l}$ (blue histograms),
$(\bar{\chi}^2)_{\ell(\scriptsize{\mbox{obs}})}^{l}$ (purple histograms)
and for the difference $(\bar{\chi}^2)_{\ell(\scriptsize{\mbox{th}})}^{l} - 
(\bar{\chi}^2)_{\ell(\scriptsize{\mbox{obs}})}^{l}$
(solid red line) for maps with mask KQ75. The final probabilities
are shown in Table (\ref{tab:prob-finais-kq75}).
\label{fig:hist-kq75}}
\end{figure}

Our objective here, for completeness, is to perform the search for planar 
signatures in foreground-cleaned masked CMB maps. For this we will compare
the masked (observational) variables $y$ with the masked (theoretical) variables 
$x$. Following the WMAP science team recommendation to reject residual 
foregrounds contamination in temperature analyses of the CMB data 
we shall use the KQ85 mask, which cuts 18.3\% of the sky data. 
Additionally, we also test the sensitivity of our results under the effect
of the more severe KQ75 mask, which represents a cut-sky of 28.4\%. Our final results 
for such analyses are shown in Tables (\ref{tab:prob-finais-kq85}) and (\ref{tab:prob-finais-kq75})
and Figs. (\ref{fig:hist-kq85}) and (\ref{fig:hist-kq75}).

Regarding the results of our analysis, we start by noting that the sectors 
$(l,\ell) = (2,5)$, $(4,7)$ and $(6,8)$, which have 
relative probabilities of 6.1\%, 6.8\% and 5\% in the full-sky maps case,
have now relative lower bounds given by 
23.1\%, 20.9\% and 11.1\% for the mask KQ85 and 43.7\%, 21.5\%, 
and 14.6\% for the mask KQ75, respectively. This reinforces an earlier claim that 
the planar anomalies detected in these scales are a consequence of residual 
foreground contaminations in the full-sky maps~\cite{Pereira:2009kg}. 

On the other hand, the angular scale $\ell=5$ at $l=10$ is still
below the 10\% threshold, even when we apply the severe KQ75 mask. Curiously, 
this scale was also reported by other groups \cite{Copi:2005ff,Eriksen:2004jg} 
as being anomalously spherical. 
The 10\% level we found here corroborates this outcome, since a very negative value 
of $(\bar{\chi}^2)^l_\ell$ indicates a lack of planar modulations.

The reader may have noticed that the sectors $(10,8)$ and $(2,9)$ in the KQ75 case 
are also below the 10\% threshold. 
We argue that these results could be showing that the foreground cleaning 
procedures (used to generate the full-sky CMB map) may not only create 
false anomalies, but also hide others, in this case the sticky points at 
the angular scales $\ell=8, 9$, which became clear only
after the application of the recommended cut-sky mask.
Nevertheless, whether these anomalies are real or caused 
by some systematic error in the map-making process, 
they suggest that further analyses are welcome. 

\begin{table}[H]
\begin{centering}
\begin{tabular}{cccccccccc}
\toprule 
$l\backslash\ell$ & \multicolumn{1}{c}{2} & \multicolumn{1}{c}{3} & 4 & 5 & 6 & 7 & 8 & 9 & 10\tabularnewline
\midrule
\midrule 
2 & 57.8\% & 60.2\% & 47.0\% & 23.1\% & 51.1\% & 60.7\% & 58.8\% & 15.5\% & 84.9\%\tabularnewline
\midrule 
4 & 81.6\% & 75.7\% & 66.1\% & 34.5\% & 52.1\% & \textcolor{black}{79.1\%} & 76.1\% & 61.9\% & 50.3\%\tabularnewline
\midrule 
6 & 75.2\% & 88.7\% & 54.3\% & 46.2\% & 31.6\% & 56.3\% & 88.9\% & 16.9\% & 88.4\%\tabularnewline
\midrule 
8 & 82.2\% & 77.6\% & 20.3\% & 43.1\% & 20.9\% & 47.0\% & 37.1\% & 29.7\% & 82.0\%\tabularnewline
\midrule 
10 & 65.2\% & 38.1\% & 52.8\% & \textbf{3.2}\% & 60.8\% & 60.4\% & 36.7\% & 32.5\% & 68.9\%\tabularnewline
\bottomrule
\end{tabular}

\caption{Final probabilities of detecting, in a random $\Lambda$CDM universe with a 
galactic cut of $18.3\%$ (KQ85 mask), a chi-square value smaller or equal to $(\bar{\chi}^2)_{\ell(\scriptsize{\rm obs})}^{l}$.
\label{tab:prob-finais-kq85}}
\par\end{centering}
\end{table}

Finally, we turn to the coherence of the angular-planar spectra $(\bar{\chi}^2)^l_\ell$ 
over a range of $l$'s and $\ell$'s. First, notice that the angular quadrupole $\ell=2$
shows again excessively positive modulations both for the KQ85 and KQ75 masks
--although for the KQ75 case the planar $l=2$ spectrum is marginally negative.
The angular octopole $\ell=3$ is now only slightly anomalous: four out of the five
spectra we computed turn out to be positive. The other case we discussed in
the full-sky maps, $l=4$, is still marginal: eight out of the nine spectra assume 
positive values.

However, the most prominent feature that appears more clearly in the cut-sky 
maps is that $\ell=5$ has an anomalous coherence, with mostly
rather negative values. In fact, $\ell=7$ and $\ell=10$ also show a
high degree of uniformity with larger-than-expected modulations. As
we showed in a previous Section, this type of coherence is the hallmark
of a preferred plane (actually, of a disk-like region around a preferred plane.) 
Although we have not produced a test for this coherent signal, we believe
our findings show that these angular scales deserve further investigation,
and, given that they appear at approximately the angular size of the
galactic cut, the coherence in $(\bar{\chi}^2)^l_\ell$ could be pointing towards
residual contaminations in the cut-sky maps.

\begin{table}[H]
\begin{centering}
\begin{tabular}{cccccccccc}
\toprule 
$l\backslash\ell$ & \multicolumn{1}{c}{2} & \multicolumn{1}{c}{3} & 4 & 5 & 6 & 7 & 8 & 9 & 10\tabularnewline
\midrule
\midrule 
2 & 47.6\% & 41.4\% & 44.3\% & 43.7\% & 51.4\% & 73.3\% & 55.9\% & \textbf{7.3}\% & 72.2\%\tabularnewline
\midrule 
4 & 78.1\% & 73.1\% & 54.5\% & 39.7\% & 50.0\% & \textcolor{black}{78.5\%} & 62.7\% & 74.5\% & 62.9\%\tabularnewline
\midrule 
6 & 60.2\% & 86.1\% & 61.2\% & 35.9\% & 25.0\% & 58.6\% & 85.4\% & 10.6\% & 75.0\%\tabularnewline
\midrule 
8 & 86.6\% & 70.5\% & 33.4\% & 52.0\% & 35.4\% & 72.7\% & 24.6\% & 64.1\% & 77.5\%\tabularnewline
\midrule 
10 & 57.5\% & 64.7\% & 43.6\% & \textbf{4.6}\% & 67.6\% & 58.8\% & \textbf{6.6}\% & 38.3\% & 52.2\%\tabularnewline
\bottomrule
\end{tabular}
\par\end{centering}

\caption{Final probabilities of detecting, in a random $\Lambda$CDM universe with a 
galactic cut of $28.4\%$ (KQ75 mask), a chi-square value smaller or equal to $(\bar{\chi}^2)_{\ell(\scriptsize{\rm obs})}^{l}$.
\label{tab:prob-finais-kq75}}
\end{table}

\section{Conclusions\label{conclusions}}

We have used a top-down approach to search for deviations of statistical isotropy 
in the WMAP data by means of the recently introduced $\chi^2$ test for
angular-planar modulations~\cite{Pereira:2009kg}.

These tools, designed to account for the presence of physical/astrophysical
planes in temperature maps, can unveil any kind of planar modulations in these
maps in a completely model-independent way. Furthermore, since improper
subtraction of foreground contamination around the galactic plane
in CMB data may lead to planar signatures, the angular-planar statistic
can be a valuable tool to test the robustness of map cleaning
procedures. Due to its generality and model independence, the angular-planar
power spectrum $\mathcal{C}^{lm}_\ell$ can be applied to any given temperature map 
--in fact, to any map whatsoever on the sphere $S^2$. 

In this work we have applied this estimator to three classes of WMAP maps, 
namely, full-sky cleaned maps and masked cleaned maps due to the application 
of the WMAP 5-year KQ85 and KQ75 masks. 
We have included in our analyses an estimation of residual foreground contamination
as well as an assessment of these types of errors in the final
probabilities we calculated.

The analysis with full-sky maps have shown that the angular quadrupole
$\ell=2$ has a consistently positive planar modulation throughout
the range $l\in[2,10]$. This systematic positivity could be an indication
of a planar signature around $\ell=2$, although we would need to
consider a wider planar range to confirm this suspicion. On the other
hand, we have found slightly anomalous values in the angular-planar
$(l,\ell)$ sectors $(2,5)$, $(10,5)$, $(4,7)$ and $(6,8)$, with
relative probabilities of 6.1\%, 8.7\%, 6.8\% and 5\%, respectively.

As argued in~\cite{Pereira:2009kg}, these anomalies are not drastically low,
and can be attributed to residual foreground contamination. Our analyses with 
masked maps confirms that this is indeed the case for the sectors
$(l,\ell)$=$(2,5)$, $(4,7)$ and $(6,8)$. While in the full-sky map analysis
the lower bound for the above probabilities was of $\sim$5\%, in the case
with the mask KQ75 this bound was raised to 14.6\%. This confirms
earlier claims that foreground contamination may be an important issue
when analyzing CMB anomalies, and that more reliable results should
always be based in masked maps analyses.

On the other hand, we have found a significant trace of anomaly in
the sector $(l,\ell)=(10,5)$. This value seems to be robust under the effect
of the two masks we used -- lower values being allowed with only 4.6\%
of chance in the case of the mask KQ75. Curiously, the angular scale
$\ell=5$ was reported to be highly spherical by other 
groups~\cite{Copi:2005ff,Eriksen:2004jg},
while we have found that it has a very low planar modulation. 
However, the multipole vectors used in that analysis
do not represent physical directions in a straightforward manner, so
the term ``spherical" in this context should be used carefully.
Nonetheless, it is interesting that this scale has conspicuously appeared
as anomalous in two rather different tests of planarity.\\

Perhaps the strongest indication of a possible anisotropy in the CMB maps
was provided by the coherence of our $\bar\chi^2$ test on the angular
scales corresponding to $\ell=5$, 7 and 10. As we showed,
this uniformity of mostly negative or positive values of $\bar\chi^2$ 
is equivalent to the signal left by a preferred plane (actually, a preferred
disk with some thickness related to $l$). This may point towards
contaminations in the masked maps which were left over from
the foreground-cleaning process, although further work will be necessary 
to confirm this suspicion.

As a final comment, we would like to mention that the approach we
used to calculate our final probabilities, namely, the convolution
of the observational and theoretical probability densities, is more
rigorous than the widely used method of just computing the area under
the theoretical probability density function. This approach generalizes
the former, and implements in a natural way the uncertainty inherent
in CMB observations, which in turn have a non-trivial impact in the
final probabilities predicted by a specific cosmological model. 

In what regards prospects for further developments, we would like
to point out that the tests used here are completely general and
have a wide range of applications, among which we can mention polarization
maps, tests of non-Gaussianity and even stacked maps of cosmic structure,
like the galaxy cluster catalog 2Mass \cite{2Mass}.

\subsection*{Acknowledgments}

We would like to thank Marcelo J. Rebou\c{c}as for useful suggestions during the 
early stages of this work. 
AB was supported by the Brazilian agency CNPq (309388/2008-2). 
Some of the results in this paper have been derived using the HEALPix 
package~\cite{Gorski:2004by}. We are also grateful for the use of the Legacy Archive 
for Microwave Background Data Analysis (LAMBDA)~\cite{Hinshaw:2008kr}. 
This work was supported by Funda\c{c}\~ao de Amparo \`a pesquisa do Estado 
de S\~ao Paulo (Fapesp) and by CNPq.

\bibliographystyle{h-physrev}
\addcontentsline{toc}{section}{\refname}
\bibliography{planar-signature}

\end{document}